\documentclass[letterpaper,titlepage,11pt]{article}
\pdfoutput=1
\usepackage{hyperref}
\usepackage{amssymb,amsmath,amsfonts}
\usepackage{epsfig}
\usepackage[mathscr]{euscript}

\def\clock{{\count0=\time
           \divide\count0 60
           \ifnum\count0<10 0\fi\the\count0
           \multiply\count0 -60 \advance\count0 \time
           :\ifnum\count0<10 0\fi \the\count0
         }}
\newcommand{\timestamp}{{\small\vbox{\hbox{\tt\jobname.tex}
\hbox{\the\day/\the\month/\the\year, \clock}}}}

\newcommand{\CF}{\mathcal{F}}
\newcommand{\CG}{\mathcal{G}}
\newcommand{\CL}{\mathcal{L}}
\newcommand{\CO}{\mathcal{O}}
\newcommand{\CN}{\mathcal{N}}

\newcommand{\CM}{\mathcal{M}}

\newcommand{\CZ}{\mathcal{Z}}

\def\PD{\mathscr{D}}

\newcommand{\gym}{g_{\rm YM}}

\newcommand{\ads}{\mbox{AdS}}

\newcommand{\nn}{\nonumber}

\newcommand{\spa}{\ , \ \ }

\newcommand{\ds}{\displaystyle}

\numberwithin{equation}{section}

\begin{document}

\begin{titlepage}

 \ \ 
 \vskip 1.8 cm

\centerline{\Huge \bf  Thermal DBI action for the D3-brane}
\vskip 0.4cm
\centerline{\Huge \bf   at weak and strong coupling}
\vskip 1.5cm

\centerline{\large {\bf Gianluca Grignani$\,^{1}$},  {\bf Troels Harmark$\,^{2}$},}
\vskip 0.2cm \centerline{\large  {\bf Andrea Marini$\,^{1}$}
and
{\bf Marta Orselli$\,^{1,2,3}$} }

\vskip 0.7cm

\begin{center}
\sl $^1$ Dipartimento di Fisica, Universit\`a di Perugia,\\
I.N.F.N. Sezione di Perugia,\\
Via Pascoli, I-06123 Perugia, Italy
\vskip 0.3cm
\sl $^2$ The Niels Bohr Institute, Copenhagen University  \\
\sl  Blegdamsvej 17, DK-2100 Copenhagen \O , Denmark
\vskip 0.3cm
\sl $^3$ Museo Storico della Fisica e Centro Studi e Ricerche Enrico Fermi \\
 Piazza del Viminale 1, I-00184 Rome, Italy 
\end{center}
\vskip 0.3cm

\centerline{\small\tt grignani@pg.infn.it, harmark@nbi.dk, }
\centerline{\small\tt andrea.marini@fisica.unipg.it, orselli@nbi.dk}

\vskip 1.3cm \centerline{\bf Abstract} \vskip 0.2cm \noindent
We study the effective action for finite-temperature D3-branes with an electromagnetic field at weak and strong coupling. We call this action the thermal DBI action. Comparing at low temperature the leading $T^4$ correction for the thermal DBI action at weak and strong coupling we find that the $3/4$ factor well-known from the AdS/CFT correspondence extends to the case of arbitrary electric and magnetic fields on the D3-brane. We investigate the reason for this by taking the decoupling limit in both the open and the closed string descriptions thus showing that the AdS/CFT correspondence extends to the case of arbitrary constant electric and magnetic fields on the D3-brane.

\end{titlepage}

\tableofcontents
\pagestyle{plain}
\setcounter{page}{1}

\section{Introduction, summary and conclusions}

The open/closed string duality in the special case of $N$ coincident D3-branes in a flat embedding of ten-dimensional Minkowski space contains an enormous amount of interesting physics. In the open string description the low energy excitations of the D3-brane is described by $\CN=4$ Super-Yang-Mills (SYM) theory with gauge group $U(N)$ while the closed string description at low energies is provided by the supergravity solution for $N$ large and the coupling $g_s N$ large, $g_s$ being the string coupling. From this setting one finds that the open/closed string duality essentially results in the celebrated AdS/CFT correspondence \cite{Maldacena:1997re,Gubser:1998bc,Witten:1998qj} when taking a certain decoupling limit that decouples the low energy excitations on the D3-branes.

One of the precursors of the AdS/CFT correspondence was the study of $N$ coincident D3-branes at low temperature \cite{Gubser:1996de}. At weak coupling $g_s N\ll 1$ they are described to leading order by $\CN=4$ SYM theory with free energy  $F = - \frac{1}{6} \pi^2 N^2 V_3 T^4$. At strong coupling $g_s N\gg 1$ and large $N$ one can instead describe them using a non-extremal supergravity solution in a near-extremal limit giving the free energy $F = - \frac{1}{8} \pi^2 N^2 V_3 T^4$ thus with the only difference to the open string description being a factor of $3/4$. The correct interpretation of this is provided by the AdS/CFT correspondence \cite{Gubser:1998nz}. Namely, the free energy of $SU(N)$ $\CN=4$ SYM theory at large $N$ takes the form $F = - f(\lambda) \frac{1}{6}\pi^2 N^2 V_3 T^4$ at any 't Hooft coupling $\lambda= \gym^2 N = 4\pi g_s N$ due to the conformality of $\CN=4$ SYM theory. At strong coupling $\lambda\gg 1$ the dual description of $\CN=4$ SYM theory is a black hole in $\ads_5\times S^5$ being the near-extremal limit of the non-extremal D3-brane supergravity solution. Hence the $3/4$ factor is the prediction of the free energy of $\CN=4$ SYM theory at strong coupling.

In this paper we are interested in studying a generalization of the above setting for the open/closed string duality for $N$ D3-branes with low temperature. Our object of study is the effective action for the $N$ D3-branes at finite temperature in a given thermal background of type IIB supergravity. We call this action the {\sl thermal DBI action} since it can be thought of as a generalization of the DBI action to finite temperature. At weak coupling $g_s N\ll 1$ one computes the thermal DBI action as the effective thermal action for the DBI action. At strong coupling $g_s N\gg 1$ the open/closed string duality reveals that the thermal DBI action can be computed from the black hole thermodynamics of $N$ coincident D3-branes probing the type IIB supergravity background in the sense of the blackfold approach (see Refs.~\cite{Emparan:2009cs,Emparan:2009at} for the blackfold approach and more specifically Refs.~\cite{Grignani:2010xm,Emparan:2011hg} for the application to D-branes). 

The main focus of this paper is the thermal DBI action for $N$ coincident D3-branes in a flat embedding in ten-dimensional Minkowski space with a Kalb-Ramond potential turned on. This corresponds to turning on an electromagnetic field on the D3-brane. We compute the thermal DBI action in this setting at low temperature both at weak coupling $g_s N\ll 1$ and at strong coupling $g_s N\gg 1$, finding
\begin{eqnarray}
\label{thermalDBI}
 I_{\rm eff} [ T, \gamma_{ab}, B_{ab} ]  &=& - NT_{\rm D3} \int d^4 \sigma \sqrt{-\det ( \gamma_{ab} + B_{ab} )} \times \nn \\ && \times \left( 1-   \frac{\sqrt{\det ( \delta^a_b + \gamma^{ac}B_{cb} )}}{(\gamma_{00}+\gamma^{ij} B_{0i}B_{0j})^2} f(4\pi g_s N) \frac{\pi^2 N T^4}{6 T_{\rm D3}} + \CO (T^8)\right) 
\end{eqnarray}
where $f(4\pi g_s N)$ takes the value $1$ for $g_s N \ll 1$ and $3/4$ for $g_s N \gg 1$. We also argue that the dependence on the coupling $g_s N$ factorize from the dependence on $\gamma_{ab}$ and $B_{ab}$ at intermediate values of the coupling for the $T^4$ term. 
Thus, the low energy fluctuations captured by the $T^4$ term has the same dependence on the electromagnetic field $B_{ab}$ at weak and strong coupling. This could seem highly surprising since the DBI theory is not a conformal theory and hence there are no immediate reasons that the coupling dependence should factorize.

We investigate the origin of the factorization of the $T^4$ term in \eqref{thermalDBI} as well. Considering the special case with $\gamma_{ab}=\eta_{ab}$ and $B_{ab}$ being constant we can write the free energy at low temperature as
\begin{equation}
\label{free_energy}
F(T,\vec{E},\vec{B}) = - f(4\pi g_s N) \frac{\pi^2}{6} V_3 N^2 T^4 \frac{1-\vec{E}^2 + \vec{B}^2 - (\vec{E}\cdot \vec{B})^2 }{(1-\vec{E}^2)^2} 
\end{equation}
where we introduced the notation $\vec{E}= (B_{01},B_{02},B_{03})$ and $\vec{B}=(B_{23},-B_{13},B_{12})$. We now employ a decoupling limit $l_s \rightarrow 0$ with $g_s$, $N$, $T$, $\vec{E}$ and $\vec{B}$ fixed while the scalar fields on the D3-branes should scale like $l_s^2$. Then at weak coupling $g_s N\ll1$ we find a finite action for the decoupled theory that corresponds to the free energy \eqref{free_energy} with $f(4\pi g_s N)=1$. While this can be computed using a one-loop correction we find that this decoupled action actually is $\CN=4$ SYM theory on a background with metric $G_{ab} = M_{ac} M_{bd} \eta^{cd}$ where we defined $M_{ab}=\eta_{ab}+B_{ab}$, with gauge coupling $\gym^2 = 4\pi g_s \sqrt{-\det M}$ and $\theta$-angle $\theta=2\pi \vec{E}\cdot \vec{B} / ( g_s \sqrt{-\det M})$. Thus, the low energy fluctuations giving the $T^4$ term in \eqref{thermalDBI} does come from a conformal theory. 

At strong coupling the $T^4$ term in \eqref{thermalDBI} and \eqref{free_energy} is found by considering two supergravity brane bound states, one with $\vec{E} \parallel \vec{B}$ and one with $\vec{E} \perp \vec{B}$. Then by rotational invariance of the action and free energy one can infer \eqref{free_energy} with $f(4\pi g_s N)=3/4$. Employing the same $l_s \rightarrow 0$ decoupling limit as at weak coupling on the two supergravity brane bound state one sees that this corresponds to taking certain near-extremal limits of the two supergravity solutions. In both cases this gives the type IIB background of the Poincar\'e patch $\ads_5$ black hole times $S^5$ but in coordinates corresponding to having the boundary metric $G_{ab} = M_{ac} M_{bd} \eta^{cd}$. Furthermore, from the dilaton one reads the gauge coupling $\gym^2 = 4\pi g_s \sqrt{-\det M}$ and from the axion field the $\theta$-angle $\theta=2\pi \vec{E}\cdot \vec{B} / ( g_s \sqrt{-\det M})$. Thus, one finds again the AdS/CFT correspondence though in a different coordinate system and with an axion/$\theta$-term turned on.

In conclusion, the reason for the factorization of the dependence on the coupling $g_s N$ in the $T^4$ term in \eqref{thermalDBI} and \eqref{free_energy} is that we can map the low energy fluctuations corresponding to the the $T^4$ term with a general $B_{ab}$ field to the low energy fluctuations for $B_{ab}=0$. 

Our results thus extends the manifestation of the AdS/CFT correspondence from the open/closed string duality on D3-branes to the case where a constant electromagnetic field is turned on on the brane. This opens up an interesting new avenue of research on the open/closed string duality and its holographic manifestations, namely the question of what happens when considering an electromagnetic field on the D3-branes that can vary along the brane. Locally, when we are at distances that are small compared to the variation of $B_{ab}$, we show in this paper that it corresponds to the AdS/CFT correspondence. However, when being at large distances, one finds a generalization of the AdS/CFT correspondence. We investigate this in a forthcoming publication \cite{forthcoming2014}.

Looking at the free energy \eqref{free_energy} it is interesting to note that if $\vec{B}=0$ or if $\vec{E} \parallel \vec{B}$ then the free energy diverges like $1/(1-\vec{E}^2)$ as $\vec{E}^2 \rightarrow 1$. Thus, we have a particular critical behavior with a certain critical exponent corresponding to the fact that one obtains non-commutative open string (NCOS) theory in this limit \cite{Seiberg:2000ms,Gopakumar:2000na}. Instead turning on $\vec{E}$ and $\vec{B}$ but keeping them non-parallel, one finds that the free energy instead diverges as $1/(1-\vec{E}^2)^2$ as $\vec{E}^2 \rightarrow 1$ thus the free energy diverges with a different critical exponent. It could be interesting to explore further the physics behind this.

Beyond studying the physics of the open/closed string duality our computation of the thermal DBI action \eqref{thermalDBI} also finds applications in using branes to probe thermal backgrounds of string theory. So far it has mostly been the DBI action that has been used to probe string theory backgrounds. This has in particular led to important results in the context of the AdS/CFT, AdS/QCD and more recently the AdS/CMT correspondences. The success of using the DBI action to describe D-brane probes of zero-temperature String Theory backgrounds have motivated the application of the DBI action as a probe of thermal backgrounds, particularly in the thermal versions of the above-mentioned holographic correspondences. However, as noted in \cite{Grignani:2010xm}, the DBI action does not accurately describe a D-brane probing a thermal background. This is because the D-brane DOFs on the brane will be heated up by the temperature of the background. Hence the effective action for D-branes in thermal backgrounds is modified. To accurately probe a thermal background with a D-brane one should therefore employ the thermal DBI action which we find in this paper both at weak and strong coupling.%
\footnote{For work on using thermal effective actions for branes to probe thermal backgrounds see \cite{Grignani:2010xm,Grignani:2011mr} for the construction of a thermal BIon solution which can be found using the strongly coupled thermal DBI action for the D3-brane, see \cite{Niarchos:2012pn,Niarchos:2013ia} for an M5-M2-brane generalization and see \cite{Grignani:2012iw} for $k$ fundamental strings probing $\ads_5\times S^5$ with a black hole corresponding to a solution of what one can call a strongly coupled thermal Nambu-Goto action. Finally, see \cite{Armas:2012bk,Armas:2013ota} for the construction of thermal Giant Gravitons probing thermal $\ads_5\times S^5$.}

This paper has the following content. Since the paper is based on studying the open/closed string duality on D3-branes we give a precise description in Section \ref{sec:setup} of the setup that can be used for our study of the thermal DBI action both from open string and closed string point of views. In particular, we turn on an electromagnetic field on $N$ coincident D3-branes by introducing a background Kalb-Ramond field. 

In Section \ref{sec:oneloopDBI} we consider the open string point of view which is valid at weak coupling $g_s N\ll 1$. This consists in analyzing the DBI action for a single D3-brane with an electromagnetic field turned on. We study the action in the above-mentioned decoupling limit $l_s \rightarrow 0$ and compare this to the action for $\CN=4$ SYM theory on curved space, thus identifying the action for the low energy fluctuations. We subsequently use this to compute the free energy and the thermal DBI action. In Appendix \ref{app:oneloop} we compute the same result directly as a one-loop correction to the DBI action.

In Section \ref{sec:dbi_strong} we consider the closed string point of view which is valid at strong coupling $g_s N\gg 1$. We explain in detail how to obtain the thermal DBI action in case of an electric field $\vec{E}$ turned on. Then we turn to the two most general cases $\vec{E} \parallel \vec{B}$ and $\vec{E} \perp \vec{B}$. The first one we obtain from a $\mbox{D3} \parallel (\mbox{F1} \parallel \mbox{D1})$ black brane bound state and the second from a $\mbox{D3} \parallel (\mbox{F1} \perp \mbox{D1})$ black brane bound state. The second bound state is a new supergravity solution in the literature, even in the extremal limit. We point out that it is related to previously known solutions including one which describes supertubes. We employ the T-duality transformations listed in Appendix \ref{app:Tduality}. 

Finally, in section \ref{sec:adscft} we take the above-mentioned decoupling limit $l_s\rightarrow 0$ on the closed string side in the form of certain near-extremal limits of the $\mbox{D3} \parallel (\mbox{F1} \parallel \mbox{D1})$ and $\mbox{D3} \parallel (\mbox{F1} \perp \mbox{D1})$ black brane bound states. In this way we show that we obtain the Poincar\'e patch black hole in $\ads_5$ times $S^5$ with the metric related by a change of coordinate to the one obtained from the D3-brane solution.

\section{Setup for computation of thermal DBI action}
\label{sec:setup}

This paper is devoted to the study of the thermal DBI action at weak and strong coupling. In this section we describe the precise setup for this study. We choose to focus on the D3-brane since this is the most interesting case in view of the connection with the AdS/CFT correspondence.%
\footnote{It could be interesting to consider other D$p$-branes as well in view of recent progress \cite{Smilga:2008bt,Wiseman:2013cda}.} Our main interest in this paper is to study the interplay between turning on a background Kalb-Ramond field which can be seen as a electromagnetic field on the D3-brane, turning on a temperature and going from weak to strong coupling. For this reason we choose for simplicity for our computations a setup in which the D3-brane world-volume has a flat embedding in the background of ten-dimensional Minkowski space. Our results can be readily generalized to the case of general embeddings and we comment briefly in Sections \ref{sec:oneloopDBI} and \ref{sec:dbi_strong} on how this generalization looks.

In our setup we consider $N$ coinciding D3-branes in the background of ten-dimensional Minkowski space with metric $\eta_{\mu\nu} dx^\mu dx^\nu$. The embedding of the D3-branes is described by $X^\mu(\sigma)$ where $\sigma^a$, $a=0,1,2,3$, are the world-volume coordinates of the D3-branes. We put the $N$ coinciding D3-branes at the hyperplane $x^4=x^5= \cdots = x^9 = 0$. Thus, we choose the following flat embedding of the D3-branes
\begin{equation}
X^a (\sigma) = \sigma^a\ , \ a=0,1,2,3\ , \ X^{i+3}(\sigma) = 0 \ , \ i =1,2,3,4,5,6
\end{equation}
Our type IIB supergravity background furthermore has zero dilaton field $\phi=0$ and zero Ramond-Ramond field strengths. We turn on a Kalb-Ramond field $B_{\mu\nu}$ in the directions parallel to the brane world-volume while being zero along the transverse directions. From the type IIB supergravity EOM's we get that the Kalb-Ramond field obeys $dB_{(2)} = 0$ in our setup. Note that for the Kalb-Ramond field $B_{(2)}$ this background is pure gauge (at least if it is topologically trivial). However, when adding D3-branes it is not pure gauge anymore, as the $B_{(2)}$ is tied to the world-volume gauge field on the D3-branes. 
 Notice also that for this background there are no forces on the D3-branes in the transverse directions hence the above is a consistent choice of embedding.

In putting the D3-branes at $x^4=x^5= \cdots = x^9 = 0$ we assume that the transverse length scale of the D3-branes $r_s$ ($i.e.$ the size in the $x^4,...,x^9$ directions) is always much smaller than the length scale $R$ over which the Kalb-Ramond field vary along the D3-branes. This means that over distances $r_s \ll r \ll R$ we can treat the brane as an infinitely thin stack of $N$ D3-branes in a space-time in which one can find world-volume coordinates such that the world-volume metric as well as the pull-back of the Kalb-Ramond field are constant.  

For the open string description we have $g_s N \ll 1$ which means we are in the weak coupling regime of the DBI action. In particular DBI describes $N=1$ with $g_s \ll 1$. In this case the DBI action can be used to describe the brane over distances of order $R$ since we have assumed $r_s \ll R$. When considering distances along the world-volume of order $r_s \ll r \ll R$ we are considering the DBI action with flat embedding in flat space with a constant Kalb-Ramond field.

For the closed string description we have $g_s N \gg 1$ which means we are in the strong coupling regime of the DBI action. In this case the stack of D3-branes backreact on the supergravity background at distance scales $r_s$ (since $r_s$ grows large when $g_s N$ does). Thus, we impose that the type IIB supergravity fields asymptote to the above chosen type IIB background for $\sqrt{ (x^4)^2 + \cdots + (x^9)^2} \rightarrow \infty$, $i.e.$ arbitrarily far away from the brane. When considering distances along the world-volume of order $r \ll R$ the closed string description is that of the supergravity background of $N$ D3-branes that asymptotically has a constant Kalb-Ramond field. Since $r_s \ll R$ we can use the blackfold approach to describe the brane system over distances of order $R$ \cite{Emparan:2009cs,Emparan:2009at}.

\section{Thermal DBI action at weak coupling and low temperature}
\label{sec:oneloopDBI}

In this section we compute the leading temperature dependent term in the effective action for D3-branes at low temperature in the case of arbitrary electric and magnetic fields on the brane at weak coupling $g_s N \ll 1$. This corresponds to computing the one-loop correction to the DBI action for D3-branes at low temperature.

We consider here the DBI action for a single D3-brane with the setup given in Section \ref{sec:setup}. Note that a single D3-brane means that $N=1$ in the notation of Section \ref{sec:setup}. This means in particular that the background Ramond-Ramond fields and dilaton field are zero. Thus, for this setup the DBI action takes the form
\begin{equation}
\label{DBIact}
I_{\rm DBI} = - T_{\rm D3} \int d^4 \sigma \sqrt{- \det ( \gamma_{ab}+B_{ab} + 2\pi l_s^2 F_{ab} -2(2\pi)^2l_s^4 \bar{\psi}\Gamma_a \partial_b \psi ) } 
\end{equation}
Note in particular that the Wess-Zumino term is absent since the Ramond-Ramond fields are set to zero. Here $\gamma_{ab}$ is the induced metric
\begin{equation}
\gamma_{ab} = \partial_a X^\mu \partial_b X^\nu \eta_{\mu\nu}
\end{equation}
and $F_{ab}$ is the $U(1)$ gauge field on the D3-brane $F_{ab} = \partial_a A_b - \partial_b A_a$.
Finally, for the fermionic part of the action \eqref{DBIact} $\psi$ is a 10-dimensional Majorana-Weyl spinor and $\Gamma_a$ are the 10-dimensional gamma matrices.%
\footnote{We rescaled the fermions with a $2\pi l_s^2$ factor for use below.}
The field strength obeys $dF = 0$ on the world-volume. The dual world-volume field strength $H_{ab}$ is the anti-symmetric world-volume tensor defined here as
\begin{equation}
\sqrt{-\gamma} H^{ab} = - \frac{1}{T_{\rm D3} 2\pi l_s^2} \, \frac{\partial \CL}{\partial F_{ab} } 
\end{equation}
The EOMs for the $U(1)$ gauge field on the D3-brane are
\begin{equation}
\label{Fequations}
\partial_a (\sqrt{-\gamma} H^{ab})= 0 \spa dF = 0
\end{equation}
where the first equation is obtained from the action by varying the gauge field $A_a$.

\subsubsection*{Classical configuration}

The classical configuration of the D3-brane is the one described in the setup of Section \ref{sec:setup}. Thus, we consider a flat embedding of the D3-brane 
\begin{equation}
\label{clasconf1}
X^a (\sigma) = \sigma^a\ , \ a=0,1,2,3 \spa X^i (\sigma) = 0 \ , \ i=4,5,...,9
\end{equation}
The induced metric is thus $\gamma_{ab} = \eta_{ab}$. For the Kalb-Ramond field we have $dB  = 0$. Furthermore, for the pullback of the Kalb-Ramond field to the world-volume we have $dB =0$ here understood as an equation on the four-dimensional world-volume.  

The goal in the following is to consider small open string fluctuations on the D3-brane in the classical closed string background described above. Therefore, for the classical configuration we impose $F=0$. Instead the Kalb-Ramond field $B_{\mu\nu}$ of the ten-dimensional background is turned on such that $(dB)_{\mu\nu\rho}=0$ with $B$ being constant in the $x^4,x^5,...,x^9$ directions transverse to the brane. One easily sees that the pullback of the Kalb-Ramond field $B_{ab} = \partial_a X^\mu \partial_b X^\nu B_{\mu\nu}$  obeys $(dB)_{abc} = 0$ as well. 
The dual field strength becomes 
\begin{equation}
\label{clastildeF}
H^{ab}= \frac{1}{2} \sqrt{-\det M} (M^{-1} )^{[ab]}
\end{equation}
with the definition 
\begin{equation}
\label{defM}
M_{ab} \equiv \eta_{ab} + B_{ab}
\end{equation}
and where $[ab]$ means antisymmetrization $(M^{-1})^{[ab]} = \frac{1}{2} ((M^{-1})^{ab}-(M^{-1})^{ba})$. Thus, the pullback of the Kalb-Ramond field $B_{ab}$ is required to obey the following Born-Infeld non-linear electromagnetic EOMs
\begin{equation}
\label{BIelectro}
 \partial_a H^{ab} = 0 \spa \partial_{[a} B_{bc]} = 0
 \end{equation}
where here $H^{ab}$ is given by Eqs.~\eqref{clastildeF}-\eqref{defM}.

\subsubsection*{Quantum fluctuations and decoupling limit}

We now consider open string quantum fluctuations on top of the chosen classical background. For the embedding this means we consider a static gauge $X^a(\sigma)=\sigma^a$, $a=0,1,2,3$, plus allowing for transverse fluctuations of $X^i(\sigma)$, $i=4,5,...,9$.
As is well-known from studying tree-level open string theory the transverse fields of the D3-brane have fluctuations that scales like $l_s^2$ where $l_s$ is the string length. Thus, we can write $X^{i}(\sigma) =2\pi l_s^2 \Phi^{i}(\sigma)$, $i=4,5,...,9$.
The induced metric then gives 
\begin{equation}
\label{metric}
\gamma_{ab} = \eta_{ab} + (2\pi)^2 l_s^4 \partial_a \Phi^i \partial_b \Phi^i
\end{equation}
where a sum over $i=4,5,...,9$ is understood.

We now consider the decoupling limit $l_s \rightarrow 0$ for a general $B_{ab}$ field. We find the following expansion of the action \eqref{DBIact} in this limit
\begin{equation}
\label{DBIact_exp}
I = I_0 + I_1 + I_2 + \CO (l_s^2)
\end{equation}
where
\begin{equation}
\label{I01}
I_0 = - T_{\rm D3} \int d^4 \sigma \sqrt{- \det M} \spa I_1 = - T_{\rm D3} 2\pi l_s^2 \int d^4 \sigma H^{ab} F_{ab}
\end{equation}
and with
\begin{equation}
\label{I2_all}
I_2 = I_{2,F} + I_{2,\Phi} + I_{2,\psi}
\end{equation}
Here we record
\begin{eqnarray}
I_{2,F} &=& - \frac{1}{2\pi g_s} \int d^4\sigma \frac{1}{4} \sqrt{-\det M} \Big[ (M^{-1})^{(ac)}(M^{-1})^{(bd)} - (M^{-1})^{[ac]}(M^{-1})^{[bd]} \nn \\[3mm] && + \frac{1}{2} (M^{-1})^{[ab]}(M^{-1})^{[cd]}     \Big] F_{ab} F_{cd}
\end{eqnarray}
\begin{equation}
I_{2,\Phi} = - \frac{1}{2\pi g_s} \int d^4\sigma \frac{1}{2} \sqrt{-\det M} (M^{-1})^{(ab)} \partial_a \Phi^i \partial_b \Phi^i
\end{equation}
\begin{equation}
\label{I2_psi}
I_{2,\psi} = \frac{1}{\pi g_s} \int d^4 \sigma \frac{1}{2} \sqrt{-\det M} (M^{-1})^{ab}  \bar{\psi}\Gamma_a \partial_b \psi
\end{equation}
where $(ab)$ means symmetrization $(M^{-1})^{(ab)} = \frac{1}{2} ( (M^{-1})^{ab}+(M^{-1})^{ba})$.

We notice from \eqref{I01} that $I_0$ is a constant depending on the given background $B_{ab}$ field. This constant scales like $l_s^{-4}$. Instead $I_1$ scales like $l_s^{-2}$ and depends on the fluctuations of the $U(1)$ gauge field $F_{ab}$. However, it follows from the Born-Infeld non-linear electromagnetic EOM  of \eqref{BIelectro} that $I_1$ is a total derivative. Hence the open string dynamics of the D3-brane in the decoupling limit $l_s \rightarrow 0$ is described by the action $I_2$ of Eqs.~\eqref{I2_all}-\eqref{I2_psi}. For $B_{ab}=0$ it is well-known that this action is $\CN=4$ super-Yang-Mills theory (SYM) with gauge group $U(1)$ and gauge coupling $\gym^2 = 4 \pi g_s$. Below we find an interpretation for the action also for $B_{ab} \neq 0$.

\subsubsection*{Interpretation of action for constant $B_{ab}$ field}

We consider now the special case of a constant $B_{ab}$ field. Using $M_{ab}$ as defined in \eqref{defM} we can define the two matrices
\begin{equation}
\label{defGE}
(G^{-1})^{ab} \equiv (M^{-1})^{(ab)} 
\spa
(E^{-1})_a {}^b \equiv \eta_{ac} (M^{-1})^{cb}
\end{equation}
where $(ab)$ means the symmetric part. These two matrices can be interpreted as an inverse metric and inverse vierbein, respectively, since we have the identity
\begin{equation}
\eta^{cd} (E^{-1})_c{}^a (E^{-1})_d{}^b = (G^{-1})^{ab}
\end{equation}
This identity follows from $(M^{-1})^{(ab)} = \eta_{cd} (M^{-1})^{ca}(M^{-1})^{db}$, or, equivalently, $M_{ac}M_{bd} (M^{-1})^{(cd)}=\eta_{ab}$.
One can easily derive the latter by noticing that it is the symmetric part of the equation $M_{ac}M_{bd} (M^{-1})^{cd}=M_{ab}$. 
From the above we can furthermore find expressions for the vierbein and metric
\begin{equation}
\label{vierbein_metric}
E_a{}^b = M_{ac} \eta^{cb} \spa G_{ab} = M_{ac} M_{bd}  \eta^{cd} 
\end{equation}
where the metric is found from $G_{ab} = \eta_{cd} E_a{}^c E_b{}^d$.
We also note that
\begin{equation}
\sqrt{-G} \equiv \sqrt{- \det G} = \det E = -\det M
\end{equation}
Using the above we see immediately
\begin{equation}
\label{new_I2_phi}
I_{2,\Phi} = - \frac{1}{2\pi g_s \sqrt{-\det M}} \int d^4\sigma \frac{1}{2} \sqrt{- G} (G^{-1})^{ab} \partial_a \Phi^i \partial_b \Phi^i
\end{equation}
\begin{equation}
\label{new_I2_psi}
I_{2,\psi} = \frac{1}{2\pi g_s \sqrt{-\det M}} \int d^4 \sigma  \sqrt{-G} (E^{-1})_a {}^b  \bar{\psi}\Gamma^a \partial_b \psi
\end{equation}
For the $F^2$ term we find
\begin{equation}
\label{new_I2_F}
I_{2,F} = - \frac{1}{2\pi g_s \sqrt{-\det M}}  \int d^4 \sigma \left\{ \frac{1}{4} \sqrt{-G}  (G^{-1})^{ac} (G^{-1})^{bd} F_{ab} F_{cd} + \frac{1}{8} \vec{E} \cdot \vec{B} \, \epsilon^{abcd} F_{ab}F_{cd} \right\}
\end{equation}
where we introduced the notation $\vec{E}= (B_{01},B_{02},B_{03})$ and $\vec{B}=(B_{23},-B_{13},B_{12})$ so that $\vec{E}\cdot \vec{B} = \frac{1}{8}\epsilon^{abcd} B_{ab}B_{cd}$. 

Consider instead the action for 
$U(1)$ $\CN=4$ SYM with a $\theta$-angle term on curved space with metric $g_{\mu\nu}$ and vierbein $e_\mu{}^a$
\begin{eqnarray}
I_{\CN=4} &=& - \frac{2}{\gym^2} \int d^4 \sigma \sqrt{-g} \left\{ \frac{1}{4} g^{\mu\rho}g^{\nu\lambda} F_{\mu\nu} F_{\rho\lambda} + \frac{1}{2} g^{\mu\nu} D_\mu \Phi^i D_\nu \Phi^i - \bar{\psi} \Gamma^a e_a {}^\mu D_\mu \psi \right\}  \nn \\ && - \frac{\theta}{32\pi^2} \int d^4 \sigma  \,  \epsilon^{abcd} F_{ab}F_{cd}
\end{eqnarray}
where $D_\mu$ is the four-dimensional covariant derivative (including spin connection for the fermions). If we choose a background with a constant metric, $i.e$ with $\partial_\mu g_{\nu\rho}=0$, then this action reduces to 
\begin{eqnarray}
I_{\CN=4} &=& - \frac{2}{\gym^2} \int d^4 \sigma \sqrt{-g} \left\{ \frac{1}{4} g^{\mu\rho}g^{\nu\lambda} F_{\mu\nu} F_{\rho\lambda} + \frac{1}{2} g^{\mu\nu} \partial_\mu \Phi^i \partial_\nu \Phi^i - \bar{\psi} \Gamma^a e_a {}^\mu \partial_\mu \psi \right\}  \nn \\ && - \frac{\theta}{32\pi^2} \int d^4 \sigma  \,  \epsilon^{abcd} F_{ab}F_{cd}
\end{eqnarray}
Comparing now this action with the action $I_2= I_{2,F} + I_{2,\Phi} + I_{2,\psi}$ given by \eqref{new_I2_phi}-\eqref{new_I2_F} we see that the action $I_2$ is the action for $\CN=4$ SYM on a background with metric $g_{\mu\nu}$ and  vierbein $e_\mu{}^{b}$ given by
\begin{equation}
\label{map_metric_B}
\begin{array}{c}
g_{\mu\nu} \leftrightarrow G_{ab} = M_{ac}  M_{bd} \eta^{cd}  \ \ \mbox{with} \ \ \mu \leftrightarrow a , \nu\leftrightarrow b
\\[3mm]
e_\mu  {}^{b} \leftrightarrow E_a{}^b = M_{ac}\eta^{cb} \ \ \mbox{with} \ \ \mu \leftrightarrow a , b \leftrightarrow b
\end{array}
\end{equation}
We should furthermore make the following identification for the gauge coupling and $\theta$ angle of $\CN=4$ SYM
\begin{equation}
\label{gym}
\gym^2 = 4 \pi g_s \sqrt{-\det M} \spa \theta =  2\pi \frac{\vec{E}\cdot \vec{B}}{g_s \sqrt{- \det M}}
\end{equation}
In conclusion, we have shown that for an arbitrary constant $B_{ab}$ field the action \eqref{I2_all}-\eqref{I2_psi} is equivalent to the action of $\CN=4$ SYM with a $\theta$ angle on a background with constant metric given in terms of $B_{ab}$ by $G_{ab}=M_{ac} M_{bd}\eta^{cd} $  and the gauge coupling and $\theta$ angle by \eqref{gym}.

We remark here that since $B_{ab}$ is constant we are considering $\CN=4$ SYM on flat space. However, for a generic $B_{ab}$ field the metric is not conformally equivalent to $\eta_{ab}$. Hence in this sense the physics is different than $\CN=4$ SYM on flat space with metric $\eta_{ab}$. However, note that $\CN=4$ SYM on any background with a constant metric has $PSU(2,2|4)$ symmetry.

\subsubsection*{Free energy at finite temperature for a constant $B_{ab}$ field}

We would like to find the one-loop correction to the thermal DBI action at weak coupling for a constant $B_{ab}$ field. This can be found explicitly by computing the free energy of the decoupled theory with action \eqref{I2_all}-\eqref{I2_psi}. However, one can also employ the equivalence of this action to $U(1)$ $\CN=4$ SYM on flat space with vierbein \eqref{map_metric_B}. The free energy of $U(1)$ $\CN=4$ SYM on flat space with metric $\eta_{ab}$ is given by 
\begin{equation}
\label{Fbefore}
F(T) = - \frac{\pi^2}{6} V_3 T^4
\end{equation}
at weak coupling $\gym^2 \ll 1$.
Consider a constant metric $g_{\mu\nu}$ as given by the map defined in \eqref{map_metric_B}. Since the metric is constant we are still in flat space and we can view the vierbein defined in \eqref{map_metric_B} as the linear map between Minkowski space in coordinates with metric $\eta_{ab}$ and coordinates with metric $g_{\mu\nu}$. Transforming from $\eta_{ab}$ to $g_{\mu\nu}$ the temperature $T$ transforms to $T /\sqrt{-g_{00}}$ where $\sqrt{-g_{00}}$ is the norm of the vector $\partial /\partial t$ (with $t$ being the time coordinate of the $g_{\mu\nu}$ system). This corresponds to the norm of $e^a{}_t$ as measured in the coordinates for $\eta_{ab}$ which in general includes both a rescaling and a boost factor. The transformation of the free energy $F(T)$ can be read off from knowing that $F(T)/T$ is equal to the Euclidean action and is thus proportional to $\sqrt{\det{g_E}}$ where $g_E$ is the Euclidean metric. Since $\det g_E = - \det g$ where $g$ is the Lorentzian signature metric we see that $F(T)/T$ transforms to $\sqrt{-g} F(\frac{T}{\sqrt{-g_{00}}})  \frac{\sqrt{-g_{00}}}{T}$ and therefore $F(T)$ transforms to $\sqrt{-g} F(\frac{T}{\sqrt{-g_{00}}})$. Hence \eqref{Fbefore} transforms to 
\begin{equation}
\label{Fgab}
F(T) = - \frac{\pi^2}{6} \sqrt{-g} V_3 \left( \frac{T}{\sqrt{-g_{00}}} \right)^4
\end{equation}
where $T$ is the temperature measured with respect to the time $t$ for $g_{\mu\nu}$ and $V_3$ is the volume measured with respect to the spatial coordinates of $g_{\mu\nu}$. To find the free energy for the decoupled theory \eqref{I2_all}-\eqref{I2_psi} one should now use the map \eqref{map_metric_B} and insert the metric as function of the Kalb-Ramond field $B_{ab}$ in the expression \eqref{Fgab}. We find 
\begin{equation}
\label{genF_gauge}
F(T,\vec{E},\vec{B}) = - \frac{\pi^2}{6} V_3 T^4 \frac{1-\vec{E}^2 + \vec{B}^2 - (\vec{E}\cdot \vec{B})^2 }{(1-\vec{E}^2)^2} 
\end{equation}
where we again used the notation $\vec{E}= (B_{01},B_{02},B_{03})$ and $\vec{B}=(B_{23},-B_{13},B_{12})$ thus viewing $B_{ab}$ as consisting of an electric part $\vec{E}$ and a magnetic part $\vec{B}$. Note that $E_i=B_{0i}$ and $B_i= \frac{1}{2} \epsilon^{ijk}B_{jk}$ with $i,j,k=1,2,3$ and $\epsilon^{123}=1$.

The free energy \eqref{genF_gauge} is the free energy in the ensemble with constant $B_{ab}$ field, $i.e.$ with constant electric $\vec{E}$ and magnetic $\vec{B}$ fields. In Section \ref{sec:dbi_strong} we shall consider the $T^4$ term for low temperature and strong coupling $g_s N \gg 1$. For this purpose it will prove useful to express the result \eqref{genF_gauge} in an ensemble independent fashion. This is done by computing the corresponding entropy
\begin{equation}
S = - \left( \frac{\partial F}{\partial T} \right)_{\vec{E},\vec{B}} 
\end{equation}
thus giving
\begin{equation}
\label{weak_entropy}
S =  \frac{2\pi^2}{3} V_3 T^3 \frac{1-\vec{E}^2 + \vec{B}^2 - (\vec{E}\cdot \vec{B})^2 }{(1-\vec{E}^2)^2} 
\end{equation}

As stated above, one can alternatively compute the free energy \eqref{genF_gauge} by performing a one-loop computation for the DBI action. This means one should expand the DBI action around the classical configuration \eqref{clasconf1}-\eqref{BIelectro} to quadratic order. From the above one sees easily that the quadratic action corresponds to the action of the decoupled theory \eqref{I2_all}-\eqref{I2_psi}. In Appendix \ref{app:oneloop} we perform this one-loop calculation in the special case of only having an electric field, so with $\vec{E}=(E,0,0)$ and $\vec{B}=0$. The result is \eqref{feintegral} which we see is in perfect correspondence with the general formula \eqref{genF_gauge}.

\subsubsection*{Thermal DBI action at finite coupling}

We note that one can argue for a generalization of the above argument for \eqref{Fgab} to include higher order terms in the coupling $g_s N$ and possibly even to finite coupling. Indeed, the generalization of \eqref{Fbefore} is \cite{Gubser:1998nz}
\begin{equation}
F(T) = - f(4\pi g_s N) \frac{\pi^2}{6}  V_3 T^4
\end{equation}
where $f(4\pi g_s N)$ is a function only of the coupling $g_s N$. That it has this form follows simply from the fact that $\CN=4$ SYM is conformal. It is known from computations in thermal $\CN=4$ SYM as well as using the holographically dual black hole in $\ads_5\times S^5$ that this function takes values \cite{Gubser:1998nz,Fotopoulos:1998es}
\begin{equation}
f(g) = \left\{ \begin{array}{l}  1 -\frac{3}{2\pi^2}g +\cdots  \ \ \mbox{for} \ \ g \ll 1 \\[2mm] \frac{3}{4} + \frac{45}{32} \zeta(3) (2g)^{-\frac{3}{2}} + \cdots \ \ \mbox{for} \ \ g \gg 1\end{array} \right.
\end{equation}
Above we argued that computing the one-loop correction to the thermal DBI action for a constant $B_{ab}$ field is equivalent to computing the free energy of $\CN=4$ SYM at finite temperature with the background metric and vierbein given by \eqref{map_metric_B}. When raising the coupling $g_s N$ in the decoupling limit $l_s \rightarrow 0$ of the DBI action this is still valid since the background given by $B_{ab}$ remains the same. Thus, this implies that we can use $\CN=4$ SYM to compute the $T^4$ correction to any order in the coupling $g_s N$ (with the Yang-Mills coupling given by \eqref{gym}). Therefore, we argue that the one-loop correction to the thermal DBI action is factorized as $F(T,\vec{E},\vec{B})= f(4\pi g_s N) F(T,\vec{E},\vec{B})|_{g_s N=0}$ and hence we have
\begin{equation}
\label{genF_gauge2}
F(T,\vec{E},\vec{B}) = - f(4\pi g_s N) \frac{\pi^2}{6} V_3 T^4 \frac{1-\vec{E}^2 + \vec{B}^2 - (\vec{E}\cdot \vec{B})^2 }{(1-\vec{E}^2)^2} 
\end{equation}
This implies in particular that we have a prediction for the strong coupling limit $g_s N \gg 1$  namely that we have a factor $3/4$ in comparison to the weak coupling result \eqref{genF_gauge2}. We shall see below that this indeed is confirmed by explicit computation at strong coupling.

\subsubsection*{Thermal DBI action including one-loop contribution}

We find the following effective action for a single D3-brane with a general embedding $X^\mu(\sigma)$ in a static thermal background with local temperature $T$, world-volume metric $\gamma_{ab} = \partial_a X^\mu \partial_b X^\nu g_{\mu\nu}$ and pullback of the Kalb-Ramond field $B_{ab} =\partial_a X^\mu \partial_b X^\nu B_{\mu\nu}$
\begin{equation}
\label{weak_TDBI}
I_{\rm eff} = - T_{\rm D3} \int d^4 \sigma \sqrt{-\det ( \gamma_{ab} + B_{ab} )}  \left( 1-  \frac{\sqrt{\det ( \delta^a_b + \gamma^{ac}B_{cb} )}}{(\gamma_{00}+\gamma^{ij} B_{0i}B_{0j})^2}  \frac{\pi^2 T^4}{6 T_{\rm D3}} + \CO (T^8)\right) 
\end{equation}
This effective action - the thermal DBI action for a D3-brane - is valid for a general type IIB supergravity background with static metric $g_{\mu\nu}$ and Kalb-Ramond field $B_{\mu\nu}$, but with constant dilaton field and zero Ramond-Ramond fields. Note also that it requires working in a static gauge such that $\gamma_{0i}=0$.

Note that we have not included a world-volume gauge field since this is an effective action for the D3-brane for a given thermal type IIB supergravity background and hence we have integrated out the degrees of freedom living on the brane such as the world-volume gauge field, as well as the scalars and the fermions corresponding to other modes of fluctuations of the D3-brane. Presumably one can trivially generalize this action to a varying dilaton background by making the substitution $T_{\rm D3} \rightarrow T_{\rm D3} e^{-\phi}$. Including the coupling to Ramond-Ramond fields would consist in adding the standard topological Wess-Zumino term to the action.

In Appendix \ref{app:oneloop} we argue that the generalization of the above effective action to the effective action of a single D$p$-brane in a static thermal background - the thermal DBI action -  is
\begin{equation}
\label{weak_TDBI_p}
\begin{split}
I_{\rm eff} = - T_{\rm Dp} &\int d^{p+1} \sigma \sqrt{-\det ( \gamma_{ab} + B_{ab} )} \ \ \times \\
 &\left( 1 -  \frac{\sqrt{\det ( \delta^a_b + \gamma^{ac} B_{cb} )}}{(-\gamma_{00}-\gamma^{ij} B_{0i}B_{0j})^{\frac{p+1}{2}}} \frac{(p-1)!\, \zeta_H\left(p+1,\frac{1}{2}\right)}{4^{p-2}\pi^{p/2}\Gamma\left(\frac{p}{2}\right)} \frac{ T^{p+1}}{ T_{\rm Dp}} + \CO (T^{2p+2})\right) \, ,
\end{split}
\end{equation}
where $\zeta_H$ is the Hurwitz zeta function defined as $\zeta_H(s,a)=\displaystyle\sum_{n=0}^\infty(n+a)^{-s}$.

\section{Thermal DBI action at strong coupling}
\label{sec:dbi_strong}

In this section we compute the thermal DBI action for D3-branes at strong coupling $g_s N \gg 1$. This is done for finite values of the temperature $T$ in the cases of parallel and orthogonal electric and magnetic fields, $i.e.$ $\vec{E} \parallel \vec{B}$ and $\vec{E} \perp \vec{B}$, respectively.
The framework for computing the thermal DBI action, which is the effective action for $N$ coincident D3-branes in a thermal background of type IIB supergravity, is the blackfold approach \cite{Emparan:2009cs,Emparan:2009at}. Using the blackfold approach to infer the effective D-brane action was first suggested in \cite{Grignani:2010xm}. 

We begin in Section \ref{sec:sugra_solutions} with finding the relevant solutions of type IIB supergravity that can be used as input in finding the strong coupling thermal DBI action. This involves two brane bound states, a $\mbox{D3} \parallel (\mbox{F1} \parallel \mbox{D1})$ and a $\mbox{D3} \parallel (\mbox{F1} \perp \mbox{D1})$ black brane bound state. The latter solution is new. It is similar in nature to the black supertube solutions \cite{Mateos:2001qs,Emparan:2001ux,Elvang:2003mj,Elvang:2004xi}.

In Section \ref{sec:strong_DBI_electric} we consider how to find the thermal DBI action with an electric field at strong coupling. We go through this case first since it is simpler and hence can be used to illustrate various important points in the procedure for the two general cases. In particular we compare to the corresponding thermal DBI action with electric field at weak coupling and we also make some general considerations on how to find the free energy corresponding to the thermal DBI action in the correct thermodynamical ensemble. 

Finally in Section \ref{sec:strong_DBI} we employ the black brane bound states found in Section \ref{sec:sugra_solutions} to get the thermal DBI action for D3-branes with two general configurations of constant electric and magnetic fields, namely the electric and magnetic fields being either parallel or orthogonal. We use this to compare to the weak coupling result for the thermal DBI action.

\subsection{Supergravity solutions for black D3-branes with $\vec{E} \parallel \vec{B}$ and $\vec{E} \perp \vec{B}$}
\label{sec:sugra_solutions}

In this section we find the solutions of type IIB supergravity describing black $\mbox{D3} \parallel (\mbox{F1} \parallel \mbox{D1})$ and $\mbox{D3} \parallel (\mbox{F1} \perp \mbox{D1})$ brane bound states. We use the following bosonic action for type IIB supergravity in the string frame
\begin{eqnarray}
I_{\rm IIB} &=& \frac{1}{2\kappa^2} \int d^{10} x \, \sqrt{-g} \left[  e^{-2\phi} \Big( R+4(\nabla \phi)^2-\frac{1}{12} H_{(3)}^2 \Big) -   \frac{1}{2} F_{(1)}^2 - \frac{1}{12} F_{(3)}^2 - \frac{1}{4 \cdot 5!} F_{(5)}^2  \right] \nn \\ &&  + \frac{1}{4\kappa^2}  \int A_{(4)} \wedge H_{(3)} \wedge d A_{(2)}
\end{eqnarray}
where $2\kappa^2= (2\pi)^7 g_s^2 l_s^8$, $g_{\mu\nu}$ is the ten-dimensional string frame metric, $\phi$ is the dilaton, $B_{\mu\nu}$ the Kalb-Ramond field, $\chi$, $A_{(2)}$ and $A_{(4)}$ are the Ramond-Ramond 0-form, 2-form and 4-form potentials, respectively, and 
we have the Kalb-Ramond and Ramond-Ramond field-strengths
\begin{equation}
H_{(3)} = dB_{(2)} \spa F_{(1)} = d\chi \spa F_{(3)} = dA_{(2)} - \chi H_{(3)} \spa F_{(5)} = dA_{(4)} - A_{(2)} \wedge H_{(3)}
\end{equation}
In addition to the EOMs derived from this action one should impose self-duality of the Ramond-Ramond five-form field strength $F_{(5)}^* = F_{(5)}$.

\subsubsection*{Black F1-D3 and D1-D3 brane bound states}

We begin with reviewing the black F1-D3 and D1-D3 brane bound states. Consider first the black F1-D3 brane bound state with string frame metric\cite{Harmark:2000wv}
\begin{equation}
\label{F1D3metric1}
ds^2 = \frac{1}{\sqrt{DH}} \left[ - f dt^2 + dx_1^2 + D ( dx_2^2 + dx_3^2 ) \right] + \frac{\sqrt{H}}{\sqrt{D}} \left[ f^{-1} dr^2 + r^2 d\Omega_5^2 \right]
\end{equation}
with dilaton, Kalb-Ramond field and Ramond-Ramond fields given by
\begin{equation}
\label{BF1D3}
\begin{array}{c}
e^{2\phi} = D^{-1}
\spa
B^{(2)} = \sin \theta ( H^{-1} - 1 ) \coth \alpha \, dt \wedge dx_1
\\[2mm]
A^{(2)} = \tan \theta ( H^{-1} D - 1 ) dx_2 \wedge dx_3=\sin \theta \cos\theta D( H^{-1} - 1 ) dx_2 \wedge dx_3
\\[2mm]
A^{(4)} = \cos \theta ( H^{-1} - 1 ) \coth \alpha \, dt \wedge dx_1 \wedge dx_2 \wedge dx_3
\end{array}
\end{equation}
where we defined
\begin{equation}
\label{harmHfD}
H = 1 + \frac{r_0^4 \sinh^2\alpha}{r^4 } \spa f = 1 - \frac{r_0^4}{r^4} \spa
D^{-1} = \cos^2 \theta + \sin^2 \theta H^{-1}
\end{equation}
Turning now to the black D1-D3 brane bound state we have the string frame metric
\begin{equation}\label{D1D3metric}
ds^2 = \frac{1}{\sqrt{H}} \left[ - f dt^2 + dx_1^2 + D ( dx_2^2 + dx_3^2 ) \right] + \sqrt{H} \left[ f^{-1} dr^2 + r^2 d\Omega_5^2 \right]
\end{equation}
with dilaton, Kalb-Ramond field and Ramond-Ramond fields given by
\begin{equation}\label{A2D1D3}
\begin{array}{c}
e^{2\phi} = D
\spa
B^{(2)} = \tan \theta ( H^{-1} D - 1 ) dx_2 \wedge dx_3
\\[2mm]
A^{(2)} =-\sin \theta ( H^{-1} - 1 ) \coth \alpha \, dt \wedge dx_1
\\[2mm]
A^{(4)} = \cos \theta D ( H^{-1} - 1 ) \coth \alpha \, dt \wedge dx_1 \wedge dx_2 \wedge dx_3
\end{array}
\end{equation}
along with identical definitions \eqref{harmHfD}.

\subsubsection*{Black $\mbox{D3} \parallel (\mbox{F1} \parallel \mbox{D1})$ brane bound state}

The black $\mbox{D3} \parallel (\mbox{F1} \parallel \mbox{D1})$ solution can be found by T-dualizing the F1-D3 brane bound state solution \eqref{F1D3metric1}-\eqref{harmHfD}. The general T-duality map used for this is recorded in Appendix \ref{app:Tduality}. We first perform a T-duality along $x^2$. Then we rotate the $(x^1,x^2)$ plane with angle $\varphi$. And then we T-dualize again along the rotated $x^2$ coordinate. This gives the black $\mbox{D3} \parallel (\mbox{F1} \parallel \mbox{D1})$ brane bound state solution

\begin{equation}
\label{D3F1D1parmetric}
ds^2 = \frac{1}{\sqrt{DH}} \left[ - f dt^2 + dx_1^2 + D E   ( dx_2^2 + dx_3^2 ) \right] + \frac{\sqrt{H}}{\sqrt{D}} \left[ f^{-1} dr^2 + r^2 d\Omega_5^2 \right]
\end{equation}
with dilaton, Kalb-Ramond field and Ramond-Ramond fields given by
\begin{equation}
\label{BD3F1D1par}
\begin{array}{c}
e^{2\phi} = D^{-1} E 
\\[2mm]
B^{(2)} = \sin \theta ( H^{-1} - 1 ) \coth \alpha \, dt \wedge dx_1 +\cos \varphi \sin \varphi E (D H^{-1}-1) dx_2 \wedge dx_3
\\[2mm]
\chi=-\tan \theta \sin \varphi (D H^{-1}-1)
\\[2mm]
A^{(2)} = - \cos\theta ( H^{-1}  - 1 ) \Big[\sin \varphi \coth \alpha\, dt \wedge dx_1+\sin \theta  \cos \varphi D E \, dx_2 \wedge dx_3 \Big]
\\[2mm]
A^{(4)} = \cos \theta \cos \varphi E (H^{-1}-1) \coth \alpha \, dt \wedge dx_1 \wedge dx_2 \wedge dx_3
\end{array}
\end{equation}
where $f$, $H$ and $D$ are defined in \eqref{harmHfD} and  $E$ is defined as
\begin{equation}
\label{harmE}
E^{-1} = \cos^2 \varphi+ \sin^2 \varphi D H^{-1} \, .
\end{equation}
%

\subsubsection*{Black $\mbox{D3} \parallel (\mbox{F1} \perp \mbox{D1})$ brane bound state}

To find the black $\mbox{D3} \parallel (\mbox{F1} \perp \mbox{D1})$ brane bound state we begin with a black D3-brane solution
\begin{equation}
\label{D3metric1}
ds^2 = \frac{1}{\sqrt{H}} \left[ - f dt^2 + dx_1^2 +  dx_2^2 + dx_3^2  \right] + \sqrt{H} \left[ f^{-1} dr^2 + r^2 d\Omega_5^2 \right]
\end{equation}
with dilaton $\phi=0$ and with Ramond-Ramond 4-form potential given by
\begin{equation}
\label{D3pot}
A^{(4)} = ( H^{-1} - 1 ) \coth \alpha \, dt \wedge dx_1 \wedge dx_2 \wedge dx_3
\end{equation}
where we defined
\begin{equation}
\label{harmD3}
H = 1 + \frac{r_0^4 \sinh^2\alpha}{r^4 } \spa f = 1 - \frac{r_0^4}{r^4}
\end{equation}
In the following we again use the T-duality map of Appendix \ref{app:Tduality}. We begin by T-dualizing the above black D3-brane along $x^3$ to reveal a smeared black D2-brane solution. We now rotate along the $(x^2,x^3)$ plane with angle $\varphi$. Following this we boost along the rotated $x^3$ direction with rapidity $\eta$. Finally, we T-dualize along the rotated and boosted $x^3$ direction. This gives the black $\mbox{D3} \parallel (\mbox{F1} \perp \mbox{D1})$ brane bound state solution

\begin{equation}\label{D3F1D1perpmetric}
\begin{split}
	ds^2& =\frac{\tilde{D}}{\sqrt{H}} \left[ - f\, (\cos^2 \varphi+ \sin^2 \varphi H^{-1})  dt^2 -2 f \, (H^{-1}-1)\sinh \eta \sin \varphi \cos \varphi \, dt \, dx_2	
	+\tilde{D} ^{-1} dx_1^2 
	\right. \\
	&\left. +   \left(\cosh^2\eta-\sinh^2\eta(\sin^2 \varphi+\cos^2 \varphi H^{-1}) f \right) dx_2^2 + dx_3^2  \right] + \sqrt{H} \left[ f^{-1} dr^2 + r^2 d\Omega_5^2 \right]
\end{split}
\end{equation}
with dilaton, Kalb-Ramond field and Ramond-Ramond fields given by
\begin{equation}
\label{BD3F1D1perp}
\begin{array}{c}
e^{2\phi} = \tilde{D}
\\[2mm]
B^{(2)} = \tilde{D} ( H^{-1} - 1 ) \cosh \eta \Big[\sinh \eta \left(\cos^2\varphi+\sinh^{-2}\alpha \right) dt \wedge dx_3 
+\cos \varphi \sin \varphi \, dx_2 \wedge dx_3\Big] 
\\[2mm]
A^{(2)} = ( H^{-1} -1 ) \coth \alpha \Big[ - \sin \varphi \, dt \wedge dx_1 + \cos \varphi \sinh \eta \,  dx_1 \wedge dx_2   \Big]
\\[2mm]
A^{(4)} = \tilde{D}(H^{-1}-1)\coth\alpha \cos\varphi \cosh\eta \, dt \wedge dx_1 \wedge dx_2 \wedge dx_3
\end{array}
\end{equation}
where $H$, $f$ and $\tilde{D}$ are defined as
\begin{equation}
\label{harmD3F1D1perp}
H = 1 + \frac{r_0^4 \sinh^2\alpha}{r^4 } \spa f = 1 - \frac{r_0^4}{r^4} \spa \tilde{D} =\frac{H}{\cosh ^2\eta \sin ^2\varphi+H \cosh ^2\eta \cos
   ^2\varphi -f \sinh ^2\eta }
\end{equation}

The above solution is new also in the extremal limit. However, it is connected to the type of brane bound states used for supertubes. Indeed, if we T-dualize along the direction of the electric field we get a (smeared) F1-D0-D2 brane bound state with is related to the original supertube construction of Townsend and Mateos \cite{Mateos:2001qs}. One can infer from this that it is a $1/4$ BPS state at zero temperature. Indeed, if one starts with an extremal $\mbox{D3} \parallel (\mbox{F1} \perp \mbox{D1})$ brane bound state smeared on a fourth direction $x^4$ and one T-dualize along $x^4$ and uplift to M-theory one gets a $\mbox{M5} \parallel ( \mbox{M2} \perp \mbox{M2})$ brane bound state which is a special case of the configurations considered in \cite{Elvang:2004ds}.

\subsection{Thermal DBI action with electric field at strong coupling}
\label{sec:strong_DBI_electric}

As a warm-up to the two most general cases we consider first how to find the thermal DBI for an electric field on the D3-brane. This parallels the methods of \cite{Grignani:2010xm} although the thermal DBI action was not written down in that paper. To do this we consider the black F1-D3 brane bound state solution \eqref{F1D3metric1}-\eqref{harmHfD}. This solution has charge quantization condition
\begin{equation}
\label{F1D3_chargequant}
N = 2\pi ^2 T_{\rm D3} \cos \theta r_0^4 \cosh \alpha \sinh \alpha
\end{equation}
where $N$ is the number of coincident D3-branes in the bound state. The expectation value of the electric field - $i.e.$ the electric part of the Kalb-Ramond field - is read off from \eqref{BF1D3} to be%
\footnote{Strictly speaking this is the value of the Kalb-Ramond field at asymptotic infinity since one should impose that it is zero on the horizon. However, in the solution \eqref{BF1D3} we use a gauge where $B_{\mu\nu}$ is zero at infinity and hence the expectation value should be read off as minus the value of $B_{\mu\nu}$ at the horizon.}
\begin{equation}
\label{F1D3_Efield}
E = B_{01} = \sin \theta \tanh \alpha
\end{equation}
The F-string charge $q_{\rm F1}$ and the corresponding number of F-strings $k$ in the bound state are
\begin{equation}
q_{\rm F1} = \frac{k}{2\pi l_s^2} = N T_{\rm D3} V_{23} \sin \theta r_0^4 \cosh \alpha \sinh \alpha
\end{equation}
where $V_{23}$ is the area along the $x^2$ and $x^3$ directions.
The conjugate chemical potential to $q_{\rm F1}$ is
\begin{equation}
\mu_{\rm F1} = V_1 E
\end{equation}
where $V_1$ is the length in the $x^1$ direction. The temperature $T$ and entropy $S$ are 
\begin{equation}
\label{F1D3_TS}
T = \frac{1}{\pi r_0 \cosh \alpha} \spa S = 2\pi^3 V_3 T_{\rm D3}^2 r_0^5 \cosh \alpha
\end{equation}
where $V_3=V_1V_2$ is the volume along the $x_1$, $x_2$ and $x_3$ directions.
One finds the Helmholtz free energy $\CF = M-TS$
\begin{equation}
\CF = \frac{\pi^2}{2} T_{\rm D3}^2 V_3 r_0^4 (1+4\sinh^2\alpha)
\end{equation}
with variation $d\CF = - S dT + \omega dN + \mu_{\rm F1} dq_{\rm F1}$. However, this is not the correct free energy for the strongly coupled thermal DBI action corresponding to the one we found at weak coupling in Section \ref{sec:oneloopDBI}. The thermal DBI action is in an ensemble of fixed temperature $T$ and fixed Kalb-Ramond field $B_{ab}$ (pulled back to the world-volume), $i.e.$ fixed electric and magnetic fields on the world-volume $\vec{E}$ and $\vec{B}$. In the present case of consideration it means we should work in an ensemble of fixed temperature $T$ and fixed electric field $E$ (as well as fixed D3-brane charge). Since the chemical potential $\mu_{\rm F1}$ is proportional to the electric field we can switch to the proper ensemble by considering the Gibbs free energy $\CG = \CF - \mu_{\rm F1} q_{\rm F1} = M - TS - \mu_{\rm F1} q_{\rm F1}$ 
\begin{equation}
\CG = \frac{\pi^2}{2} T_{\rm D3}^2 V_3 r_0^4 (1+4 \cos^2 \theta \sinh^2\alpha)
\end{equation}
with variation $d\CG = - S dT + \omega dN - q_{\rm F1} d\mu_{\rm F1} $. Using \eqref{F1D3_chargequant} and \eqref{F1D3_Efield} we can write this as
\begin{equation}
\label{F1D3_gibbs_x}
\CG = N T_{\rm D3} V_3 \sqrt{1-E^2} \frac{x-\frac{3}{4}}{\sqrt{x^2-x}} \spa x \equiv (1-E^2) \cosh^2\alpha
\end{equation}
Combining \eqref{F1D3_chargequant} with the temperature \eqref{F1D3_TS} we find
\begin{equation}
\frac{\pi^2 N T^4}{2 T_{\rm D3}} = \cos \theta \frac{\sinh\alpha}{\cosh^3\alpha}
\end{equation}
This gives the third order equation
\begin{equation}
\label{electric_3rd}
\frac{4 \cos^2 \delta}{27} x^3 - x + 1 =0
\end{equation}
where we defined
\begin{equation}
\cos \delta \equiv \frac{3 \sqrt{3} \pi^2 N T^4}{4 T_{\rm D3}(1-E^2)^{3/2}} 
\end{equation}
Note that \eqref{electric_3rd} only has solutions with $x$ real and non-negative if $\cos \delta \leq 1$. This gives an upper bound on the temperature
\begin{equation}
\frac{T^4}{T_{\rm D3}} \leq \frac{4\sqrt{3}}{9\pi^2 N} (1-E^2)^{3/2}
\end{equation}
Or, alternatively, an upper bound on the electric field
\begin{equation}
E \leq \sqrt{1 - \Big( \frac{9\pi^2 N T^4}{4\sqrt{3} T_{\rm D3}} \Big)^{\frac{2}{3}} }
\end{equation}
The third order equation \eqref{electric_3rd} gives in general two different physical branches, one connected to the extremal F1-D3 brane bound state (reached for $\delta=\pi/2$), corresponding to
\begin{equation}
\label{electric_xsol1}
x = \frac{3}{2} \frac{\cos \frac{\delta}{3} + \sqrt{3} \sin \frac{\delta}{3}}{\cos \delta}
\end{equation}
and another branch connected to a 3-brane made of smeared extremal F-strings (reached for $\delta=\pi/2$),
\begin{equation}
\label{electric_xsol2}
x = \frac{3}{2} \frac{\cos \frac{\delta}{3} - \sqrt{3} \sin \frac{\delta}{3}}{\cos \delta}
\end{equation}
Both are valid branches for the thermal DBI action. However, since we would like to compare to the thermal DBI action at weak coupling and low temperature as found in Section \ref{sec:oneloopDBI} the relevant branch to consider is \eqref{electric_xsol1} since this is connected to the extremal F1-D3 brane bound state which gives the strong coupling analogue of the zero temperature DBI action. Inserting \eqref{electric_xsol1} into the Gibbs free energy \eqref{F1D3_gibbs_x} now gives the thermal DBI action for a given temperature $T$, electric field $E$ and number of D3-branes $N$. For $T^4 \ll T_{\rm D3} (1-E^2)^{3/2}$ we see that one has an expansion of $\CG$ as
\begin{equation}
\CG = N T_{\rm D3} V_3 \sqrt{1-E^2} \left( 1 - \sum_{n=1}^\infty a_n \frac{T^{4n}}{T_{\rm D3}^n (1-E^2)^{3n/2}} \right)
\end{equation}
One can easily compute the first couple of coefficients to find
\begin{equation}
\label{F1D3_gibbs}
\CG = N T_{\rm D3} V_3 \sqrt{1-E^2} \left( 1 - \frac{\pi^2 N T^4}{8 T_{\rm D3} (1-E^2)^{3/2}} -  \frac{\pi^4 N^2 T^8}{32 T_{\rm D3}^2 (1-E^2)^{3}} + \CO(T^{12}) \right)
\end{equation}
This is the thermal DBI action at strong coupling for the D3-brane with an electric field at low temperature. 

\subsubsection*{Comparison to weakly coupled thermal DBI action}

If we consider the leading term in \eqref{F1D3_gibbs} this corresponds to $I_0$ in \eqref{I01} for an electric field. This matches perfectly, as a consequence of the F1-D3 brane bound state being $1/2$ BPS at zero temperature.%
\footnote{There is a factor of $N$ in the strong coupling result not present in \eqref{I01} at weak coupling. However, this factor of $N$ is trivially found by combining $N$ times the result at weak coupling since two extremal F1-D3 bound states do not have any force between them.} Looking instead at the first correction at low temperature we see that the $T^4$ term in \eqref{F1D3_gibbs} precisely is $3/4$ times the result \eqref{genF_gauge} for $\vec{E}=(E,0,0)$ and $\vec{B}=0$ at weak coupling. Note that the $N^2$ factor seen at strong coupling in \eqref{F1D3_gibbs} trivially occurs also at weak coupling when going to $N$ D3-branes since the weak coupling result is computed for the free theory and hence $N^2$ just counts the dimension of the adjoint representation of $SU(N)$.%
\footnote{The dimension of the adjoint representation of $SU(N)$ is $N^2-1$ but since we assume that $N$ is large the leading order answer is $N^2$.}
Thus, we can conclude that the $3/4$ factor found without electric field turned on generalizes to the case with electric field. 

At first this could seem highly surprising since the $3/4$ factor is due to the conformality of the theory on the D3-brane for small excitations since they are governed by the $\CN=4$ SYM theory and also since it is a consequence of the AdS/CFT correspondence. However, we have already shown in Section \ref{sec:oneloopDBI} that we can map small fluctuations of the F1-D3 brane at weak coupling to $\CN=4$ SYM in a different coordinate system and we shall see in Section \ref{sec:adscft} that at strong coupling we can analogously get a map to the black hole in $\ads_5\times S^5$ from the decoupling limit of the F1-D3 brane bound state. Thus, in this sense the $3/4$ factor for a general electric field is not surprising but a consequence of the fact that the behavior of small excitations of D3-branes with constant electric and magnetic fields can be mapped to the behavior of small fluctuations of the D3-brane without electric and magnetic fields. Indeed, we shall see in Section \ref{sec:strong_DBI} that the $3/4$ factor persists for more general electric and magnetic field configurations as well. Note that it is important in this that the map from excitations of a D3-brane with electric and magnetic field to the ones without electric and magnetic fields is independent of the coupling $g_s N$. This is indeed what we argued in Section \ref{sec:oneloopDBI} where it is seen that the dependence of the coupling and the dependence of the electric and magnetic fields factorize in \eqref{genF_gauge2}. In support of this we find that the same coordinate transformation defined by \eqref{map_metric_B} is found for $g_s N \gg 1$ in terms of the black hole in $\ads_5\times S^5$ in Section \ref{sec:adscft}.

\subsubsection*{Comparison to weakly coupled thermal DBI action without choosing ensemble}

In the above we succesfully found the thermal DBI action with electric field at strong coupling $g_s N \gg 1$. In doing this, it seemed important to find the free energy in the right ensemble, namely the one with fixed electric field. This is possible since the electric field is proportional to the chemical potential conjugate to the F-string charge in the F1-D3 brane bound state. However, this poses a problem for more general configuration of electric and magnetic field. If we consider the case of a D3-brane with a magnetic field it is described at strong coupling as the D1-D3 brane bound state \eqref{D1D3metric}, \eqref{A2D1D3} and \eqref{harmHfD}. Here the Helmholtz free energy $\CF=M-TS$ is in the ensemble of fixed $T$, $N$ and D1-brane charge $q_{\rm D1}$. Alternatively the Gibbs free energy $\CG=M-TS - \mu_{\rm D1} q_{\rm D1}$ is in the ensemble of fixed $T$, $N$ and D1-brane chemical potential $\mu_{\rm D1}$. However, none of these ensembles correspond to the ensemble of fixed magnetic field. This can be seen by comparing the magnetic field $B$ to $\mu_{\rm D1}$ and $q_{\rm D1}$,
\begin{equation}
B = \frac{\tan \theta \sinh^2 \alpha}{\cosh^2 \alpha + \tan^2 \theta} \spa \mu_{\rm D1} = V_1 \sin \theta \tanh \alpha \spa q_{\rm D1} = N T_{\rm D3} V_{23} \sin \theta r_0^4 \cosh \alpha \sinh \alpha
\end{equation}
Hence, rather than struggling to find the appropriate ensemble it would be better to avoid having to use a particular ensemble to find the thermal DBI action. This is possible by employing the entropy instead of the free energy. The entropy is the same regardless of ensembles, the choice of ensembles only comes in as a change of variables. Hence the strategy below in the cases with more complicated electric and magnetic field configurations will be to find the entropy $S$ and expressing it as a function of $T$, $N$, $\vec{E}$ and $\vec{B}$. With this in hand, one can integrate to find the free energy in the ensemble of fixed $T$, $N$, $\vec{E}$ and $\vec{B}$ as
\begin{equation}
\label{completeF}
F(T,N,\vec{E},\vec{B}) = N T_{\rm D3} V_3 \sqrt{1 - \vec{E}^2 + \vec{B}^2 - (\vec{E}\cdot \vec{B})^2 } - \int_0^T dT' S(T',N,\vec{E},\vec{B})
\end{equation}
where we assumed we are in the branch connected to the extremal bound state of the D3-brane with electric and magnetic fields, and we infered the leading term of the free energy from the mass of the extremal bound state.

Employing the entropy instead of the free energy to compute the thermal DBI action at strong coupling and low temperature has the additional advantage that the leading order term in the entropy corresponds to the first order term in the free energy, and the 1st order term in the entropy to the 2nd order term in the free energy, and so on. Hence to find the analogue of the $T^4$ term in \eqref{F1D3_gibbs} for the case of the electric field we only need the leading order term of the entropy. Solving \eqref{electric_3rd} to leading order simply means ignoring the constant term and hence 
\begin{equation}
x = \frac{3\sqrt{3}}{2\cos \delta} = \frac{2 T_{\rm D3} (1-E^2)^{3/2}}{\pi^2 N T^4 }
\end{equation}
to leading order (for the branch connected to the extremal F1-D3 bound state). This should be inserted in the general expression of the entropy \eqref{F1D3_TS} in terms of $T$, $E$ and $x$
\begin{equation}
S = \frac{2 T_{\rm D3}^2 V_3 (1-E^2)^2}{\pi^2 T^5 x^2}
\end{equation}
thus giving the leading order entropy
\begin{equation}
S = \frac{\pi^2}{2} N^2 V_3 T^3 \frac{1}{1-E^2} + \CO ( T^7)
\end{equation}
which we see is $3/4$ times the weak coupling result \eqref{weak_entropy} (in addition to the $N^2$ factor). Inserting this in the general formula \eqref{completeF} then reveals the leading and first order terms of \eqref{F1D3_gibbs}.

\subsection{Thermal DBI at strong coupling for $\vec{E} \parallel \vec{B}$ and for $\vec{E} \perp \vec{B}$}
\label{sec:strong_DBI}

We consider here the thermal DBI action at strong coupling for the two most general configurations of electric and magnetic fields for which we have available supergravity solutions, namely for $\vec{E} \parallel \vec{B}$ and for $\vec{E} \perp \vec{B}$.

\subsubsection*{Thermal DBI for $\vec{E} \parallel \vec{B}$}

For $\vec{E} \parallel \vec{B}$ we use the type IIB supergravity solution given by Eqs.~\eqref{D3F1D1parmetric}-\eqref{harmE} and Eq.~\eqref{harmHfD}. We have the charge quantization of the number $N$ of D3-branes
\begin{equation}
\label{EpB_N}
N = 2\pi^2 T_{\rm D3} r_0^4 \cos \varphi \cos \theta \cosh \alpha \sinh \alpha
\end{equation}
We find the following temperature and entropy
\begin{equation}
\label{EpB_TS}
T = \frac{1}{\pi r_0 \cosh \alpha} \spa S = 2\pi^3 V_3 T_{\rm D3}^2 r_0^5 \cosh \alpha = \frac{2T_{\rm D3}^2 V_3}{\pi^2 T^5 \cosh^4 \alpha}
\end{equation}
Moreover, we can read off the parallel electric and magnetic fields from the $B_{01}$ and $B_{23}$ components of the Kalb-Ramond fields \eqref{BD3F1D1par} as
\begin{equation}
\label{EpB_EB}
E = \sin \theta \tanh \alpha \spa B = \frac{\tan \varphi \cos^2 \theta \sinh^2 \alpha}{\cos^2 \theta \cosh^2 \alpha + \sin^2 \theta + \tan^2 \varphi}
\end{equation}

Combining \eqref{EpB_N} with the temperature $T$ in \eqref{EpB_TS} we find
\begin{equation}
\frac{4 \cos^2 \delta}{27} \cosh^6 \alpha- \cosh^2 \alpha+1 = 0 \spa \cos \delta \equiv \frac{3\sqrt{3} \pi^2 N T^4}{4 T_{\rm D3} \cos \varphi \cos \theta}
\end{equation}
Solving this as a third order equation for $\cosh^2\alpha$ we find the two possible physical branches
\begin{equation}
\cosh^2 \alpha = \left\{ \begin{array}{l} \ds \frac{3}{2} \frac{\cos \frac{\delta}{3} + \sqrt{3} \sin \frac{\delta}{3}}{\cos \delta} \ \ \mbox{(Connected to extremal bound state)}\\[4mm] \ds \frac{3}{2} \frac{\cos \frac{\delta}{3} - \sqrt{3} \sin \frac{\delta}{3}}{\cos \delta} \ \ \mbox{(Connected to neutral 3-brane)}\end{array} \right.
\end{equation}
The first branch is connected to the extremal $\mbox{D3} \parallel (\mbox{F1} \parallel \mbox{D1})$ brane bound state and the second to the neutral 3-brane. Combining this solution for $\cosh^2 \alpha$ with \eqref{EpB_EB} one can find the entropy $S$ as a function of $T$, $N$, $E$ and $B$ for the two branches. Using then \eqref{completeF} one finds the thermal DBI action at strong coupling. We are particularly interested in the first branch since we would like to compare to the thermal DBI at weak coupling \eqref{genF_gauge}. We can solve this explicitly for small $T$ giving
\begin{equation}
\label{EpB_entropy}
S = \frac{\pi^2}{2} V_3 N^2 T^3 \frac{1+B^2}{1-E^2}
\left(1 + \frac{\pi^2 N \sqrt{1+B^2}\,(1+2B^2)}{2T_{\rm D3}(1-E^2)^{3/2}} T^4\right) + \CO (T^{11})
\end{equation}
Here we only listed the two first contributions at low temperature but one can find any number of higher order terms in powers of the temperature from the implicit solution above.
We can compare the first term to the entropy \eqref{weak_entropy} at weak coupling for $\vec{E}\parallel \vec{B}$
\begin{equation}
S = \frac{2\pi^2}{3} V_3 T^3 \frac{1+B^2}{1-E^2} 
\end{equation}
As expected from the discussion in Section \ref{sec:strong_DBI_electric} the strongly coupled thermal DBI action has the same dependence on the electric and magnetic fields and the only difference is a factor of $3/4$ as well as the $N^2$ factor from having multiple D3-branes.

\subsubsection*{Thermal DBI for $\vec{E} \perp \vec{B}$}

For $\vec{E} \perp \vec{B}$ we use the type IIB supergravity solution given by Eqs.~\eqref{D3F1D1perpmetric}-\eqref{harmD3F1D1perp}. We have the charge quantization of the number $N$ of D3-branes
\begin{equation}
\label{EoB_N}
N = 2\pi^2 T_{\rm D3} r_0^4 \cos \varphi \cosh \eta \cosh \alpha \sinh \alpha
\end{equation}
We find the following temperature and entropy
\begin{equation}
\label{EoB_TS}
T = \frac{1}{\pi r_0 \cosh \alpha \cosh \eta} \spa S = 2\pi^3 V_3 T_{\rm D3}^2 r_0^5 \cosh \alpha \cosh \eta= \frac{2T_{\rm D3}^2 V_3}{\pi^2 T^5 \cosh^4 \alpha \cosh^4 \eta}
\end{equation}
We can read off the orthogonal electric and magnetic fields from the $B_{03}$ and $B_{23}$ components of the Kalb-Ramond fields \eqref{BD3F1D1perp} as
\begin{equation}
\label{EoB_EB}
E = \tanh \eta \spa B = \frac{\cos \varphi \sin \varphi  \sinh^2 \alpha}{\cosh \eta (\sin^2 \varphi +  \cos^2 \varphi \cosh^2\alpha)}
\end{equation}

Combining \eqref{EoB_N} with the temperature $T$ in \eqref{EoB_TS} we find
\begin{equation}
\frac{4 \cos^2 \delta}{27} \cosh^6 \alpha- \cosh^2 \alpha+1 = 0 \spa \cos \delta \equiv \frac{3\sqrt{3} \pi^2 N T^4 \cosh^3 \eta}{4 T_{\rm D3} \cos \varphi}
\end{equation}
Solving this as a third order equation for $\cosh^2\alpha$ we find the two possible physical branches
\begin{equation}
\cosh^2 \alpha = \left\{ \begin{array}{l} \ds \frac{3}{2} \frac{\cos \frac{\delta}{3} + \sqrt{3} \sin \frac{\delta}{3}}{\cos \delta} \ \ \mbox{(Connected to extremal bound state)}\\[4mm] \ds \frac{3}{2} \frac{\cos \frac{\delta}{3} - \sqrt{3} \sin \frac{\delta}{3}}{\cos \delta} \ \ \mbox{(Connected to neutral 3-brane)}\end{array} \right.
\end{equation}
The first branch is connected to the extremal $\mbox{D3} \parallel (\mbox{F1} \perp \mbox{D1})$ brane bound state and the second to the neutral 3-brane. Combining this solution for $\cosh^2 \alpha$ with \eqref{EoB_EB} one can find the entropy $S$ as a function of $T$, $N$, $E$ and $B$ for the two branches. Using then \eqref{completeF} one finds the thermal DBI action at strong coupling. We are particularly interested in the first branch since we would like to compare to the thermal DBI at weak coupling \eqref{genF_gauge}. We can solve this explicitly for small $T$ giving
\begin{equation}
\label{EoB_entropy}
S = \frac{\pi^2}{2} V_3 N^2 T^3 \frac{1-E^2+B^2}{(1-E^2)^2} 
\left(1 + \frac{\pi^2 N \sqrt{1-E^2+B^2}\,(1-E^2+2B^2)}{2T_{\rm D3}(1-E^2)^{3}} T^4\right) + \CO (T^{11})
\end{equation}
Here we only listed the two first contributions at low temperature but one can find any number of higher order terms in powers of the temperature from the implicit solution above.
We can compare the first term to the entropy \eqref{weak_entropy} at weak coupling for $\vec{E}\perp \vec{B}$
\begin{equation}
S = \frac{2\pi^2}{3} V_3 T^3 \frac{1-E^2+B^2}{(1-E^2)^2} 
\end{equation}
As expected from the discussion in Section \ref{sec:strong_DBI_electric} the strongly coupled thermal DBI action has the same dependence on the electric and magnetic fields and the only difference is a factor of $3/4$ as well as the $N^2$ factor from having multiple D3-branes.

\subsection{Thermal DBI at strong coupling and low temperature}
\label{sec:strong_DBI_lowT}

In Section \ref{sec:strong_DBI} we found explicit expressions for the thermal DBI action at low temperature and strong coupling $g_s N \gg 1$ in the cases $\vec{E} \parallel \vec{B}$ and $\vec{E} \perp \vec{B}$. These are given by \eqref{EpB_entropy} and \eqref{EoB_entropy} from which one easily finds the thermal DBI action from \eqref{completeF}. While no supergravity solution is known for general configurations of constant $\vec{E}$ and $\vec{B}$ fields one can use the two cases to infer the general thermal DBI action. This relies on the fact that the thermal DBI action transform as a scalar under rotation of the three spatial directions. Hence it can only be formulated in terms of rotational invariants. Thus, in addition to the parameters $V_3$, $N$ and $T$ the only other rotational invariants in our setup are $\vec{E}^2$, $\vec{B}^2$ and $\vec{E} \cdot \vec{B}$ (note that $(\vec{E}\times \vec{B})^2$ is not an independent invariant). Hence we can see directly from the expressions for the entropy \eqref{EpB_entropy} and \eqref{EoB_entropy} that the unique generalization to arbitrary constant $\vec{E}$ and $\vec{B}$ is
\begin{eqnarray}
\label{strong_entropy}
S &=& \frac{\pi^2}{2} V_3 N^2 T^3 \frac{1-\vec{E}^2+\vec{B}^2 - (\vec{E}\cdot \vec{B})^2}{(1-\vec{E}^2)^2} 
\nn \\ && + \frac{\pi^4 }{4T_{\rm D3}}V_3 N^3  T^7 \frac{(1-\vec{E}^2+\vec{B}^2 - (\vec{E}\cdot \vec{B})^2)^{\frac{3}{2}}}{(1-\vec{E}^2)^5}  (1-\vec{E}^2+2\vec{B}^2- 2 (\vec{E}\cdot \vec{B})^2) \nn \\ && + \CO (T^{11})
\end{eqnarray}
Inserting this into \eqref{completeF} we get the thermal DBI action 
\begin{eqnarray}
\label{strong_free_energy}
F & = & N T_{\rm D3} V_3 \sqrt{1-\vec{E}^2 + \vec{B}^2 - (\vec{E}\cdot \vec{B})^2} \left( 1 -  \frac{\sqrt{1-\vec{E}^2 + \vec{B}^2 - (\vec{E}\cdot \vec{B})^2 }}{(1-\vec{E}^2)^2} \frac{\pi^2 N T^4}{8 T_{\rm D3}} \right. \nn \\ && \left. - \frac{(1-\vec{E}^2+\vec{B}^2 - (\vec{E}\cdot \vec{B})^2)}{(1-\vec{E}^2)^5}  (1-\vec{E}^2+2\vec{B}^2- 2 (\vec{E}\cdot \vec{B})^2) \frac{\pi^4 N^2 T^8 }{32 T_{\rm D3}^2 } + \CO (T^{12})  \right)
\end{eqnarray}
in our setup of Section \ref{sec:setup}. We see that this correctly reduces to the free energy \eqref{F1D3_gibbs} for zero magnetic field $\vec{B}=0$.

One can easily generalize the thermal DBI action corresponding to the free energy \eqref{strong_free_energy} to work for a general embedding $X^\mu(\sigma)$ in a static thermal background with local temperature $T$, world-volume metric $\gamma_{ab} = \partial_a X^\mu \partial_b X^\nu g_{\mu\nu}$ and pullback of the Kalb-Ramond field $B_{ab} =\partial_a X^\mu \partial_b X^\nu B_{\mu\nu}$
\begin{equation}
\label{strong_TDBI}
I_{\rm eff} = - N T_{\rm D3} \int d^4 \sigma \sqrt{-\det ( \gamma_{ab} + B_{ab} )}  \left( 1-  \frac{\sqrt{\det ( \delta^a_b + \gamma^{ac}B_{cb} )}}{(\gamma_{00}+\gamma^{ij} B_{0i}B_{0j})^2}  \frac{\pi^2 N T^4}{8 T_{\rm D3}}  + \CO (T^{8})\right) 
\end{equation}
This effective action - the thermal DBI action for $N$ coincident D3-branes at low temperatures and strong coupling $g_s N\gg 1$ - is valid for a general type IIB supergravity background with static metric $g_{\mu\nu}$ and Kalb-Ramond field $B_{\mu\nu}$, but with constant dilaton field and zero Ramond-Ramond fields. We observe again the $3/4$ factor in comparing it to its weak coupling $g_s N \ll 1$ counterpart \eqref{weak_TDBI} (for $N=1$). 
Note that the above action requires working in a static gauge such that $\gamma_{0i}=0$. Presumably one can trivially generalize this action to a varying dilaton background by making the substitution $T_{\rm D3} \rightarrow T_{\rm D3} e^{-\phi}$. Including the coupling to Ramond-Ramond fields would consist in adding the standard topological Wess-Zumino term to the action. Note finally that one can readily generate any number of higher order corrections in powers of the temperature from the above, starting with the $T^8$ term already computed in \eqref{strong_entropy}. 

The interpretation of the action \eqref{strong_TDBI} as the action of $N$ coincident black D3-branes in a background Kalb-Ramond field at strong coupling $g_s N\gg 1$ (and with $N\gg 1$) is due to the blackfold approach (see Refs.~\cite{Emparan:2009cs,Emparan:2009at} for the blackfold approach and more specifically Refs.~\cite{Grignani:2010xm,Emparan:2011hg} for the application to D-branes).  This is valid as long as the length scales of the variations of the embedding, the metric and the Kalb-Ramond field are large compared to the charge radius of the brane $(N/T_{\rm D3})^{1/4}$.

\section{Decoupling limit and the AdS/CFT correspondence}
\label{sec:adscft}

In this section we complete the description of our understanding of the small excitations of the DBI action at weak and strong coupling in the form of the $T^4$ term at low temperature in the thermal DBI action. We do this by taking the same decoupling limit at strong coupling $g_s N \gg 1$ that we already performed on the DBI action at weak coupling $g_s N \ll 1$ in Section \ref{sec:oneloopDBI}.

\subsubsection*{Decoupling limit of $\mbox{D3} \parallel (\mbox{F1} \parallel \mbox{D1})$ brane bound state}

Consider the black $\mbox{D3} \parallel (\mbox{F1} \parallel \mbox{D1})$ brane bound state solution of type IIB supergravity given by Eqs.~\eqref{D3F1D1parmetric}-\eqref{harmE} and Eq.~\eqref{harmHfD} with properties \eqref{EpB_N}-\eqref{EpB_EB}. The decoupling limit of the DBI action at weak coupling in Section \ref{sec:oneloopDBI} translates for this supergravity solution to the limit
\begin{equation}
\label{EpB_declim}
l_s \rightarrow 0 \spa g_s, N, T, \vec{E}, \vec{B} \ \ \mbox{fixed} \spa u \equiv \frac{r}{l_s^2} \ \ \mbox{fixed}\spa u_0 \equiv \frac{r_0}{l_s^2} \ \ \mbox{fixed}
\end{equation}
Note that demanding $r$ and $r_0$ to be of order $l_s^2$ is analogous to the statement at weak coupling that transverse fluctuations of $X(\sigma)$ are of order $l_s^2$. 
We consider here the case of constant $E$ and $B$ fields. Taking now the decoupling limit \eqref{EpB_declim} we get the supergravity solution 
with metric
\begin{equation}
\label{EpB_ads}
l_s^{-2}ds^2 = \frac{u^2}{\sqrt{\lambda}} \Big[ (1-E^2) (- f dt^2 + dx_1^2 ) + (1+B^2) (dx_2^2 + dx_3^2) \Big] + \sqrt{\lambda} \left[ \frac{du^2}{u^2} +  d\Omega_5^2 \right]
\end{equation}
with
\begin{equation}
f = 1-\frac{u_0^4}{u^4}
\end{equation}
and with constant dilaton and axion fields
\begin{equation}
\label{EpB_phi_chi}
e^{2\phi} = (1-E^2)(1+B^2) \spa \chi = \frac{EB}{\sqrt{(1-E^2)(1+B^2)}}
\end{equation}
where we defined
\begin{equation}
\label{EpB_lambda}
\lambda = 4 \pi g_s e^{\phi} N =  4\pi g_s N \sqrt{(1-E^2)(1+B^2)}
\end{equation}
and with the Ramond-Ramond five-form field strength having $N$ units of flux on the $(t,x_1,x_2,x_3,u)$ part as well as on the $S^5$ part. In addition, one gets constant values for the Kalb-Ramond field $B_{(2)}$ and the Ramond-Ramond 2-form potential $A_{(2)}$. However, these can trivially be gauged away.

We recognize the solution \eqref{EpB_ads}-\eqref{EpB_lambda} as the black hole in $\ads_5\times S^5$ in the Poincar\'e patch, just in slightly different coordinates. Indeed, if we make the coordinate rescaling
\begin{equation}
\label{EpB_transf}
\tilde{t} = \sqrt{1-E^2} t \spa \tilde{x}_1 = \sqrt{1-E^2} x_1 \spa \tilde{x}_2 = \sqrt{1+B^2} x_2 \spa \tilde{x}_3 = \sqrt{1+B^2} x_3
\end{equation}
we see that we transform to the metric 
\begin{equation}
\label{usualads}
l_s^{-2}ds^2 = \frac{u^2}{\sqrt{\lambda}} \Big[ - f d\tilde{t}^2 + d\tilde{x}_1^2+ d\tilde{x}_2^2+ d\tilde{x}_3^2  \Big] + \sqrt{\lambda} \left[ \frac{du^2}{u^2} +  d\Omega_5^2 \right]
\end{equation}
This also gives an alternative derivation of the leading term in the entropy \eqref{EpB_entropy} as this can be obtained from the entropy in the $(\tilde{t},\tilde{x}_1,\tilde{x}_2,\tilde{x}_3)$ coordinates being $S= (\pi^2/2) N^2 \tilde{V}_3 \tilde{T}^3$ and then one can find \eqref{EpB_entropy} by applying the transformation \eqref{EpB_transf}.

Finally, if we take the zero-temperature limit of the solution \eqref{EpB_ads}-\eqref{EpB_lambda} (corresponding to $u_0=0$) we can compare the metric on the world-volume coordinates $(t,x_1,x_2,x_3)$ to the metric $G_{ab}$ at weak coupling as given by the map \eqref{map_metric_B}. For $B_{01}=E$ and $B_{23}=B$ the map \eqref{map_metric_B} gives
\begin{equation}
\label{EpB_wvmet}
-G_{00}=G_{11} = 1- E^2 \spa G_{22}=G_{33} = 1+B^2
\end{equation}
which we indeed recognize as the metric on the world-volume coordinates $(t,x_1,x_2,x_3)$. Thus, the constant background metric for $\CN=4$ SYM theory that we found at weak coupling $g_s N\ll 1$ also appears at strong coupling $g_s N\gg 1$. Furthermore, we see that the identification of the $\CN=4$ SYM gauge coupling in \eqref{gym} is in perfect correspondence with the identification of $\lambda$ in \eqref{EpB_lambda} as the 't Hooft coupling $\lambda = \gym^2 N$ while the $\theta$ angle in \eqref{gym} is in correspondence to the axion field $\chi$ in \eqref{EpB_phi_chi} via the holographic dictionary \cite{Banks:1998nr}
\begin{equation}
\label{chi_theta_dict}
\chi = g_s \frac{\theta}{2\pi}
\end{equation}

We can therefore conclude that starting with a D3-brane with constant and parallel electric and magnetic components of the Kalb-Ramond field turned on we get the AdS/CFT correspondence between $\CN=4$ SYM theory and $\ads_5\times S^5$ in the Poincar\'e patch with different coordinates along the world-volume directions corresponding to having the world-volume metric \eqref{EpB_wvmet} as well as a rescaled 't Hooft coupling \eqref{EpB_lambda}. A similar conclusion was reached in \cite{Lu:1999uv}. This is evidence that the dependence of the coupling constant and the electromagnetic fields in the $T^4$ term in the free energy $F(T,\vec{E},\vec{B})$ indeed factorizes at all values of the coupling as we speculated in Section \ref{sec:oneloopDBI}.

\subsubsection*{Decoupling limit of $\mbox{D3} \parallel (\mbox{F1} \perp \mbox{D1})$ brane bound state}

We now turn to the black $\mbox{D3} \parallel (\mbox{F1} \perp \mbox{D1})$ brane bound state solution of type IIB supergravity  given by Eqs.~\eqref{D3F1D1perpmetric}-\eqref{harmD3F1D1perp} with properties \eqref{EoB_N}-\eqref{EoB_EB}. The decoupling limit of the DBI action at weak coupling in Section \ref{sec:oneloopDBI} again corresponds to the limit \eqref{EpB_declim}. Taking the decoupling limit we find
\begin{eqnarray}
\label{EoB_ads}
l_s^{-2}ds^2 &=& \frac{u^2}{\sqrt{\lambda}} \left[ -  (1-E^2) f \left( dt - \frac{EB}{1-E^2} dx_2 \right)^2 + dx_1^2  +   \frac{1-E^2+B^2}{1-E^2}  dx_2^2 \right. \nn \\[3mm] && + (1-E^2+B^2) dx_3^2 \Big] + \sqrt{\lambda} \left[ \frac{du^2}{u^2} +  d\Omega_5^2 \right]
\end{eqnarray}
with
\begin{equation}
f = 1-\frac{u_0^4}{u^4}
\end{equation}
and with constant dilaton and axion fields
\begin{equation}
\label{EoB_phi_chi}
e^{2\phi} = 1 - E^2 + B^2  \spa \chi = 0
\end{equation}
where we defined
\begin{equation}
\label{EoB_lambda}
\lambda = 4 \pi g_s e^{\phi} N =  4\pi g_s N \sqrt{1-E^2+B^2}
\end{equation}
and with the Ramond-Ramond five-form field strength having $N$ units of flux on the $(t,x_1,x_2,x_3,u)$ part as well as on the $S^5$ part. In addition, one gets constant values for the Kalb-Ramond field $B_{(2)}$ and the Ramond-Ramond 2-form potential $A_{(2)}$. However, these can trivially be gauged away.

We recognize the solution \eqref{EoB_ads}-\eqref{EoB_lambda} as the black hole in $\ads_5\times S^5$ in the Poincar\'e patch, just in slightly different coordinates. Indeed, if we make the coordinate transformation
\begin{equation}
\label{EoB_transf}
\tilde{t} = \sqrt{1-E^2} \left(  t  - \frac{EB}{1-E^2} x_2 \right) \spa \tilde{x}_1 = x_1 \spa \tilde{x}_2 = \frac{\sqrt{1-E^2+B^2}}{\sqrt{1-E^2}} x_2 \spa \tilde{x}_3 = \sqrt{1-E^2+B^2} x_3
\end{equation}
we see that we transform to the metric \eqref{usualads}. This gives an alternative derivation of the leading term in the entropy \eqref{EoB_entropy} as this can be obtained from the entropy in the $(\tilde{t},\tilde{x}_1,\tilde{x}_2,\tilde{x}_3)$ coordinates and then one can find \eqref{EoB_entropy} by applying the transformation \eqref{EoB_transf}.

Finally, we take the zero-temperature limit of the solution \eqref{EoB_ads}-\eqref{EoB_lambda} (corresponding to $u_0=0$) giving the metric
\begin{eqnarray}
\label{EoB_ads_extr}
l_s^{-2}ds^2 &=& \frac{u^2}{\sqrt{\lambda}} \Big[ -  (1-E^2)  dt^2 + 2EB \, dt \, dx_2 + dx_1^2  +   (1+B^2)  dx_2^2  + (1-E^2+B^2) dx_3^2 \Big] \nn \\ && + \sqrt{\lambda} \left[ \frac{du^2}{u^2} +  d\Omega_5^2 \right]
\end{eqnarray}
We now compare the metric on the world-volume coordinates $(t,x_1,x_2,x_3)$ to the metric $G_{ab}$ at weak coupling as given by the map \eqref{map_metric_B}. For $B_{03}=E$ and $B_{23}=B$ the map \eqref{map_metric_B} gives
\begin{equation}
\label{EoB_wvmet}
G_{00} = -1+E^2 \spa G_{02} = EB \spa G_{11}=1 \spa G_{22} = 1+B^2 \spa G_{33} = 1-E^2 +B^2
\end{equation}
which we indeed recognize as the metric on the world-volume coordinates $(t,x_1,x_2,x_3)$ of \eqref{EoB_ads_extr}. Thus, the constant background metric for $\CN=4$ SYM theory that we found at weak coupling $g_s N\ll 1$ also appears at strong coupling $g_s N\gg 1$. Furthermore, we see that the identification of the $\CN=4$ SYM coupling in \eqref{gym} is in perfect correspondence with the identification of $\lambda$ in \eqref{EoB_lambda} as the 't Hooft coupling $\lambda = \gym^2 N$. Note also that $\chi=0$ in \eqref{EoB_phi_chi} is in correspondence with having $\theta=0$ in $\CN=4$ SYM theory from \eqref{gym}.

We conclude therefore that starting with a D3-brane with constant and orthogonal electric and magnetic components of the Kalb-Ramond field turned on we get the AdS/CFT correspondence between $\CN=4$ SYM theory and $\ads_5\times S^5$ in the Poincar\'e patch with different coordinates along the world-volume directions corresponding to having the world-volume metric \eqref{EoB_wvmet} as well as a rescaled 't Hooft coupling \eqref{EoB_lambda}. This gives further evidence that the dependence on the coupling constant and the electromagnetic fields in the $T^4$ term in the free energy $F(T,\vec{E},\vec{B})$ indeed factorizes at all values of the coupling as we speculated in Section \ref{sec:oneloopDBI}.

\section*{Acknowledgments}

We thank Jan de Boer, Roberto Emparan and Niels Obers for useful discussions. TH acknowledge support from the ERC-advance grant ``Exploring the Quantum Universe" as well as from the Marie-Curie-CIG grant ``Quantum Mechanical Nature of Black Holes" both from the European Union.
MO thanks the Niels Bohr Institute for hospitality.

\begin{appendix}

\section{Explicit one-loop computation of free energy in electric case}
\label{app:oneloop}

The one-loop correction to the thermal DBI action at weak coupling for a constant $B_{ab}$ field
can be derived explicitly by computing the free energy of the decoupled theory with action \eqref{I2_all}-\eqref{I2_psi}.
To do this, we first have to compute the partition function $\CZ$ which is defined as
\begin{equation}\label{oneloopZ}
	\CZ=\int \PD A_a\,\PD \Phi^i\,\PD \bar{\psi}\PD \psi\,\delta(h)\det\left(\frac{\delta h}{\delta \chi}\right)
	e^{-I_{2,F}^{\rm E}-I_{2,\Phi}^{\rm E}-I_{2,\psi}^{\rm E}}\, ,
\end{equation}
where we Wick rotated to Euclidean time. The time direction is periodic with period equal to the inverse temperature $\beta=1/T$ and we impose periodic boundary condition for the scalars and the gauge field and anti-periodic boundary conditions for the fermions. 
$h$ is a functional of the gauge field and its derivatives and it is taken to be zero,
$h\left[A,\partial A\right]=0$, in order to fix the gauge.
$\chi$ is the gauge transformation parameter.
We choose $h=A_3=0$ so that in the following we have $ \frac{\delta h}{\delta \chi}=\partial_3$. 

From above we see that the partition function factorizes in the product of the three contributions 
\begin{equation}
\CZ=\left[\int \PD A_a\, \delta(h)\det\left(\frac{\delta h}{\delta \chi}\right)e^{-I_{2,F}^{\rm E}}\right]
	\left[\int \PD \Phi^i\,e^{-I_{2,\Phi}^{\rm E}}\right]\left[\int \PD \bar{\psi} \PD \psi\, e^{-I_{2,\psi}^{\rm E}}\right]
=\CZ_{F}\CZ_{\Phi}\CZ_{\psi}\, .
\end{equation}

For the sake of simplicity, we show explicitly the computation of the partition function when only the $B_{01}$ 
component of the Kalb-Ramond field is turned on. This corresponds the pure electric case, $\vec{E}=(E,0,0)$, with
$E=B_{01}$.

Let us start with the contribution coming from the gauge field. We can write
\begin{equation}
	I_{2,F}^{\rm E}=\frac{T_{\rm D3}}{2\left(1-E^2\right)^{5/6}}
	\int_0^\beta d\tau \int d^3r \mathbf{A}^T \CM \mathbf{A}\, ,
\end{equation}
where $d^3r=dx_1dx_2dx_3$, $\mathbf{A}=\left(A_0, A_1, A_2\right)^T$ and the matrix $\CM$ is 
\begin{align}
	&\CM=	\frac{1}{\left(1-E^2\right)^{2/3}} \times \nonumber \\
	&\begin{pmatrix}
	-\partial_1^2-\left(\partial_2^2+\partial_3^2\right)\left(1-E^2\right) & \partial_0\partial_1 &
	\partial_0\partial_2 \left(1-E^2\right)\\
	\partial_0\partial_1 & -\partial_0^2-\left(\partial_2^2+\partial_3^2\right)\left(1-E^2\right) &
	 \partial_1\partial_2\left(1-E^2\right)\\
	 \partial_0\partial_2 \left(1-E^2\right) & \partial_1\partial_2 \left(1-E^2\right) &
	-\left(1-E^2\right)\left(\partial_0^2+\partial_1^2+\partial_3^2\left(1-E^2\right)\right)
\end{pmatrix}\, .
\end{align}
The partition function is then given by
\begin{equation}
\label{za}
	\CZ_F=\left(\det\partial_3\right)\left(\det \CM\right)^{-1/2}\, .
\end{equation}
The determinant of $\CM$ can be computed by substituting each entry in the matrix by
the corresponding eigenvalue and then by taking the infinite product over all the possible
eigenvalues, namely
\begin{equation}
\label{detMandf}
\det \CM =\prod_{\vec{k}}\prod_{n=-\infty}^\infty k_3^2
	\left[\left(\frac{2\pi n}{\beta}\right)^2+f(\vec{k})\right]^2 
\end{equation}
where
\begin{equation}
f(\vec{k})=k_1^2+\left(1-E^2\right)\left(k_2^2+k_3^2\right)\, .
\label{fk}
\end{equation}
Using this in \eqref{za} we find
\begin{equation}\label{Zgauge}
	\CZ_F=\prod_{\vec{k}}\prod_{n=-\infty}^\infty
	\left[\left(\frac{2\pi n}{\beta}\right)^2+f(\vec{k})\right]^{-1} .
\end{equation}

Now we take into account the contribution coming from the scalars and the fermions. The
scalar partition function is
\begin{equation}\label{Zscalars}
	\CZ_\Phi
	=\prod_{\vec{k}}\prod_{n=-\infty}^\infty
	\left[\left(\frac{2\pi n}{\beta}\right)^2+f(\vec{k})\right]^{-3} \, ,
\end{equation}
and the fermionic one is
\begin{equation}
\label{Zfermions}
	\CZ_\psi
	=\prod_{\vec{k}}\prod_{n=-\infty}^\infty
	\left[\left(\frac{\pi (2n+1)}{\beta}\right)^2+f(\vec{k})\right]^{4} \, .
\end{equation}

Putting together the above results \eqref{Zgauge}, \eqref{Zscalars}, \eqref{Zfermions}, we can compute the total free energy $F=-\frac{1}{\beta}\log \CZ$ and we obtain
\begin{equation}
\label{feintegral}
	F(T,E)=\frac{8 V_3}{\beta}\int \frac{d^3k}{(2\pi)^3}
	\left[\log \left(1-e^{-\beta \sqrt{f(\vec{k})}}\right)-\log \left(1+e^{-\beta \sqrt{f(\vec{k})}}\right)\right]=-\frac{T^4\pi^2 V_3}{6\left(1-E^2\right)}\, ,
\end{equation}
where we made use of the $\zeta$-function regularization prescription to compute the infinite product over $n$ appearing in the partition function. One can easily generalize the above computation to a general $B_{ab}$ field. This gives indeed the general result \eqref{genF_gauge} as we computed in an alternative fashion in Section \ref{sec:oneloopDBI}.

It is also interesting to further generalize this result for a D$p$-brane. Sticking again to the
pure electric case, in which the only non-vanishing component of $B_{ab}$ is $B_{01}=E$, we get 
\begin{equation}
\label{feintegralDp}
	F_{p}(T,E)=\frac{8 V_p}{\beta}\int \frac{d^pk}{(2\pi)^p}
	\left[\log \left(1-e^{-\beta \sqrt{f_p(\vec{k})}}\right)-\log \left(1+e^{-\beta \sqrt{f_p(\vec{k})}}\right)\right]\, ,
\end{equation}
where
\begin{equation}
f_p(\vec{k})=k_1^2+\left(1-E^2\right)\left(k_2^2+\cdots +k_p^2\right) \, .
\label{fkDp}
\end{equation}
In order to compute the integral \eqref{feintegralDp} it is convenient to perform the following rescaling,
$k_i \to (1-E^2)^{-1/2} k_i$ for $i=2,\dots,p$, in such a way to recover the integral for the free energy 
with zero $B_{ab}$ field. This yields
\begin{equation}
\label{feintegralDp2}
\begin{split}
	F_{p}(T,E)&=\frac{8 V_p}{\beta\left(1-E^2\right)^{\frac{p-1}{2}}}\int \frac{d^pk}{(2\pi)^p}
	\left[\log \left(1-e^{-\beta k}\right)-\log \left(1+e^{-\beta k}\right)\right] \\
	&=\frac{(p-1)!\, \zeta_H\left(p+1,\frac{1}{2}\right)T^{p+1}}{4^{p-2}\pi^{p/2}\Gamma\left(\frac{p}{2}\right)\left(1-E^2\right)^{\frac{p-1}{2}}}	
	=	\frac{F_p(T,0) }{\left(1-E^2\right)^{\frac{p-1}{2}}}\, ,
\end{split}		
\end{equation}
in which $k=\sqrt{k_1^2+\cdots+k_p^2}$ and $\zeta_H$ is the Hurwitz zeta function. One can easily generalize the above to an arbitrary constant $B_{ab}$ field giving the same result as in \eqref{weak_TDBI_p}.

\section{T-duality map}
\label{app:Tduality}

The ten-dimensional T-duality map between the type IIA and the type IIB supergravity fields was given in \cite{Bergshoeff:1995as} (see page 30-31). Using our notation and conventions the map from the type IIB to the type IIA supergravity for a T-duality along a direction $x$ reads
\begin{subequations}\label{TdualityIIBtoIIA}
\begin{align}
&\tilde{g}_{xx}=\frac{1}{g_{xx}}&  &\hspace{-3cm}e^{2\tilde{\phi}}=\frac{e^{2\phi}}{g_{xx}}\label{BA1}\\
&\tilde{g}_{\mu\nu} = g_{\mu\nu} - \frac{1}{g_{xx}} \Big(  g_{x \mu} g_{x \nu} - B_{x \mu} B_{x \nu} \Big)&
&\hspace{-3cm}\tilde{g}_{x \mu} =-\frac{B_{x \mu}}{g_{xx}}\label{BA2}\\
&\tilde{B}_{\mu\nu} = B_{\mu\nu} + \frac{1}{g_{xx}} \Big( B_{x\mu} g_{x\nu} - B_{x\nu} g_{x\mu} \Big)&
 &\hspace{-3cm}\tilde{B}_{x\mu} =- \frac{g_{x\mu}}{g_{xx}} \label{BA3}\\
&A^{(1)}_\mu=A^{(2)}_{x\mu}-\chi B_{x\mu}&  &\hspace{-3cm}A^{(1)}_x=\chi\label{BA4}\\
&A^{(3)}_{x\mu\nu}=A^{(2)}_{\mu\nu}+\frac{1}{g_{xx}} \Big(A^{(2)}_{x\mu} g_{x\nu} - A^{(2)}_{x\nu}g_{x\mu}  \Big)&&\label{BA5} \\
&A^{(3)}_{\mu\nu\rho}=A^{(4)}_{x\mu\nu\rho}+{3}\Big(A^{(2)}_{x[\mu} B_{\nu\rho]} - B_{x[\mu} A^{(2)}_{\nu\rho]} -\frac{B_{x[\mu}A^{(2)}_{|x|\nu}g_{\rho]x}-A^{(2)}_{x[\mu}B_{|x|\nu}g_{\rho]x}}{g_{xx}}\Big)&&\label{BA6}
\end{align}
\end{subequations}
The map from the type IIA to the type IIB supergravity for a T-duality along a direction $x$ reads
\begin{subequations}\label{TdualityIIAtoIIB}
\begin{align}
&\tilde{g}_{xx}=\frac{1}{g_{xx}}&    &e^{2\tilde{\phi}}=\frac{e^{2\phi}}{g_{xx}}\label{AB1}\\
&\tilde{g}_{\mu\nu} = g_{\mu\nu} - \frac{1}{g_{xx}} \Big(  g_{x \mu} g_{x \nu} - B_{x \mu} B_{x \nu} \Big)&
&\tilde{g}_{x \mu} =-\frac{B_{x \mu}}{g_{xx}}\label{AB2}\\
&\tilde{B}_{\mu\nu} = B_{\mu\nu} - \frac{1}{g_{xx}} \Big( B_{x\mu} g_{x\nu} - B_{x\nu} g_{x\mu} \Big)&
&\tilde{B}_{x\mu} =-\frac{g_{x\mu}}{g_{xx}}\label{AB3}\\
&A^{(2)}_{\mu\nu}=A^{(3)}_{x\mu\nu}+A^{(1)}_x B_{\mu\nu}+2A^{(1)}_{[\mu} B_{\nu] x}-2\frac{A^{(1)}_x}{g_{xx}} g_{x[\mu}B_{\nu] x}&    &A^{(2)}_{x\mu}= A^{(1)}_\mu- \frac{A^{(1)}_x g_{x\mu}}{g_{xx}}\label{AB4}\\
&A^{(4)}_{x\mu\nu\rho}=A^{(3)}_{\mu\nu\rho}+{3}\Big(A^{(1)}_{[\mu} B_{\nu\rho]}  -\frac{g_{x[\mu}B_{\nu\rho]}A^{(1)}_{x}}{g_{xx}}-
\frac{g_{x[\mu}A^{(3)}_{\nu\rho]x}}{g_{xx}}\Big)& &\chi=A^{(1)}_x\label{AB5}
\end{align}
\end{subequations}

\end{appendix}

\providecommand{\href}[2]{#2}\begingroup\raggedright\endgroup

\end{document}